\documentclass[12pt]{article}
\usepackage{amsfonts,graphicx,amssymb,amsmath,amsthm,epsfig}
\usepackage{float}
\allowdisplaybreaks

\begin{document}
\title{\bf Complexity for Dynamical Anisotropic Sphere in $f(G,T)$ Gravity}
\author{M. Sharif \thanks{msharif.math@pu.edu.pk} and K. Hassan \thanks{komalhassan3@gmail.com}\\
Department of Mathematics, University of the Punjab,\\
Quaid-e-Azam Campus, Lahore-54590, Pakistan.}

\date{}

\maketitle
\begin{abstract}
This paper is devoted to the formulation of a complexity factor for
dynamical anisotropic sphere in the framework of $f(G,T)$ gravity,
where $G$ is the Gauss-Bonnet invariant and $T$ is the trace of
energy-momentum tensor. Inhomogeneous energy density, anisotropic
pressure, heat dissipation and modified terms create complexity
within the self-gravitating system. We evaluate the structure
scalars by orthogonal splitting of the Riemann tensor to evaluate a
complexity factor which incorporates all the fundamental properties
of the system. Moreover, we examine the dynamics of the sphere by
assuming homologous mode as the simplest pattern of evolution. We
also discuss dissipative as well as non-dissipative scenarios
corresponding to homologous and complexity free conditions. Finally,
we establish a criterion under which the complexity free condition
remains stable throughout the process of evolution. We conclude that
the presence of dark source terms of $f(G,T)$ gravity increase the
system's complexity.
\end{abstract}
{\bf Keywords:} $f(G,T)$ gravity; Self-gravitating systems;
Complexity factor\\
{\bf PACS:} 98.62.Gq; 04.40.Dg;04.40.-b

\section{Introduction}

The universe includes small as well as large scale structures
ranging from small planets to galaxies that contain trillions of
stars. These cosmic systems influence the evolution of cosmos and
lay the foundation of cosmological study. In this regard, Einstein
presented general theory of relativity (GR) to study the dynamical
relationship among matter, gravity, space and time. This theory
opened new avenues in the field of cosmology and astrophysics by
assuming that the universe is static, i.e., it is neither
contracting nor expanding. Later in 1929, Edwin Hubble proposed that
the universe is in a phase of accelerated expansion due to the
presence of some mysterious force known as dark energy. Different
astrophysical phenomena such as Supernovae Ia and cosmic microwave
background radiation \cite{1a} confirm the accelerated expansion of
the universe. Dark matter is another mysterious component of the
universe whose presence is indicated by the dynamics of galaxies in
cluster and rotation curves of spiral galaxies. Due to the problems
of cosmic coincidence and fine tuning related to the cosmological
constant in GR, the current cosmological accelerated expansion
cannot be adequately discussed in this theory. Therefore, modified
theories are considered to explain these mysterious components. For
this purpose, the Einstein-Hilbert (EH) action is modified either by
replacing or adding the generic scalar invariants instead of
curvature scalar ($R$).

In this regard, Nojiri and Odintsov \cite{4} put forward $f(G)$
gravity by adding the generic Guass-Bonnet (GB) term in EH action.
They discussed the transition from deceleration to acceleration
phase via the Guass-Bonnet invariant. This invariant is composed of
Ricci scalar, Ricci tensor $(R_{\alpha\beta})$ and Riemann tensor
$(R_{\alpha\beta\mu\nu})$ as
$G=R^2-4R^{\alpha\beta}R_{\alpha\beta}+R^{\alpha\beta\mu\nu}R_{\alpha\beta\mu\nu}$.
Different cosmological and astrophysical phenomena have been
investigated in this theory. Bamba et al. \cite{5} determined the
stability of a bouncing cosmos by reconstructing $f(G)$ gravity.
They also studied the early-time bounce as well as late-time
acceleration of the cosmos through a scale factor involving a
combination of two exponential functions. Abbas et al. \cite{5a}
employed the Krori-Barua metric along with power-law model to
analyze the formation of compact stars in Guass-Bonnet gravity and
investigated their viable and stable behavior. Shamir and Saeed
\cite{6} examined the power as well as exponential law solutions for
static spacetime by considering a viable $f(G)$ gravity model.
Sharif and Saba \cite{6a} examined the viability and stability of
anisotropic spherical object by using decoupling technique with
suitable $f(G)$ gravity models. Sharif and Ramzan \cite{6b} adopted
embedding class-1 technique to inspect different physical aspects of
anisotropic structures.

Recently, Sharif and Ikram \cite{7} introduced a modification of
$f(G)$ with non-minimal coupling by including the trace of
energy-momentum tensor $(T)$ in the EH action which led to $f(G,T)$
gravity. They established the energy conditions for FRW universe by
reconstructing de Sitter and power-law models. It was noted that an
extra force compelled the massive particles to follow a non-geodesic
trajectory for non-zero pressure. Furthermore, they \cite{7a}
analyzed stability of the Einstein universe by using the
perturbation scheme. Hossienkhani et al. \cite{8} derived the energy
conditions for anisotropic universe filled with perfect fluid and
found that weak and null energy conditions are fulfilled for a
certain range of parameters. Moreover, they observed that an
increment in anisotropy led to the increasing behavior of weak
energy condition. Sharif and Gul \cite{17c} determined the collapse
rate of spherical object filled with perfect fluid through dynamical
equations in the context of $f(G,T)$ gravity. Yousaf \cite{9}
determined the structure scalars and examined their effects on Weyl
tensor, shear scalar and Raychaudhuri equations in the same theory.

Massive objects such as galaxies, stars and their clusters compose
most of the visible universe. The prominent characteristics
(pressure, energy density, heat flow, etc.) of the cosmic objects
are interrelated which contribute to the complex nature of these
stellar configurations. Thus, a factor mathematically relating the
aforementioned factors through a single relation aids in the study
of stellar configurations. There have been several attempts in
various scientific fields to accurately explain complexity.
Nevertheless, an interpretation of complexity which is suitable for
all fields has not been achieved. L{\'o}pez-Ruiz et al. \cite{10a}
were among the first to characterize complexity in the form of
entropy (measure of disorderness) and information. Initially, ideal
gas and perfect crystal were taken into account to define
complexity. The entropy of a perfect crystal is zero due to
symmetric arrangement of atoms. On the other hand, random motion of
particles within ideal gas generate maximum entropy. Moreover, a
small portion of perfect crystal contains characteristics of the
entire structure. Thus, its probability distribution depends on the
perfect symmetry of all accessible states whereas ideal gas has
maximum information for a brief part as all the accessible states
have same probability. However, both setups are assigned zero
complexity in physics. It means that the measure of complexity of a
system in terms of entropy and information has not proven to be
helpful.

The concept of complexity was extended to disequilibrium to check
how different probabilistic states vary from the equiprobable
distribution of the system \cite{10b}. Moreover, the complexity of
perfect crystal as well as ideal gas reduces to zero under this
definition. This definition has also been employed to calculate
complexity of some compact cosmic objects by substituting energy
density in place of probability distribution \cite{11a}. However,
this definition is not sufficient to measure the complexity as it
does not include the effects of other state determinants such as
temperature, heat flux, pressure, etc. Recently, Herrera \cite{12}
resolved this issue by identifying complexity in terms of
anisotropic pressure, inhomogeneous energy density and active
gravitational mass for a static self-gravitating matter
distribution. Four structure scalars were generated through the
orthogonal splitting of the Riemann tensor. The structure scalar
encompassing the effects of all state determinants was termed as the
complexity factor. Herrera \cite{13} extended the definition of
complexity to non-static dissipative fluid and also discussed two
patterns of evolution. This definition of complexity has also been
applied to axially symmetric spacetime \cite{14}.

Using Herrera's technique, Sharif and Butt \cite{15} determined the
complexity factor for a static matter distribution in cylindrical
spacetime. They also studied the influence of electromagnetic field
on static spherical \cite{16} as well as cylindrical \cite{17}
structures and observed an increment in complexity in the presence
of charge. The concept of complexity has also been analyzed in the
framework of $f(G)$ and $f(G,T)$ theories. Sharif et al. \cite{17a}
determined the complexity of static anisotropic spherical objects in
$f(G)$ gravity which led to more complex system. Complexity factors
have been derived for different symmetries and matter distributions
in the background of other modified theories as well \cite{18}.
Recently, Yousaf et al. \cite{17b} checked the behavior of
complexity factor in the context of $f(G,T)$ gravity for charged as
well as uncharged static sphere and concluded that complexity
increases in both scenarios. It is interesting to mention here that
the definition of complexity given in \cite{12} is not associated to
entropy and information, instead it is based on the premise that the
simplest system is characterized by the isotropic pressure in
homogeneous matter distribution.

In this paper, we evaluate the complexity factor for a non-static
dissipative spherical distribution and discuss the evolution
patterns in the framework of $f(G,T)$ gravity. We have used the
metric signatures $(-,+,+,+)$ and relativistic units, i.e., $G=c=1$.
The paper is organized as follows. In section \textbf{2}, the field
equations and essential characteristics of the matter source
describing the dynamics of the system are evaluated. In section
\textbf{3}, we decompose the Riemann tensor to obtain scalar
functions. Two patterns of evolution (homologous and homogeneous)
are discussed in section \textbf{4}. In section \textbf{5},
kinematical and dynamical quantities are derived to discuss possible
solutions for both non-dissipative and dissipative scenarios. The
deviation of self-gravitating object from zero complexity condition
is checked in section \textbf{6}. In section \textbf{7}, we
summarize the obtained results.

\section{$f(G,T)$ Gravity and Physical Variables}

The action integral for this theory is given by
\begin{equation}\label{1}
\mathbb{S}_{f(G,T)}=\frac{1}{2k^2}\int\sqrt{-g}d^{4}x[f(G,T)+R]
+\int\sqrt{-g}d^{4}x\pounds_{m},
\end{equation}
where $\pounds_{m}$, $g$ and $k^{2}$ denote the Lagrangian density,
determinant of the metric tensor ($g_{\alpha\beta}$) and coupling
constant, respectively. To proceed we consider $\pounds_{m}=P$. The
Lagrangian density and the energy-momentum tensor are related as
\begin{equation}\label{1a}
T_{\alpha\beta}=g_{\alpha\beta}\pounds_m-\frac{2\partial\pounds_m}{\partial
g^{\alpha\beta}}.
\end{equation}
The field equations are derived by varying Eq.\eqref{1} with respect
to $g_{\alpha\beta}$ as
\begin{eqnarray}\nonumber
G_{\alpha\beta}&=&8\pi
T_{\alpha\beta}-(\Theta_{\alpha\beta}+T_{\alpha\beta})f_{T}(G,T)
+\frac{1}{2}g_{\alpha\beta}f(G,T)+(4R_{\mu\beta}R^{\mu}_{\alpha}-2RR_{\alpha\beta}\\\nonumber
&+&4R^{\mu\nu}R_{\alpha\mu \beta \nu}-2R^{\mu \nu \gamma}
_{\alpha}R_{\beta \mu \nu
\gamma})f_{G}(G,T)+(4R_{\alpha\beta}\nabla^{2}+4g_{\alpha\beta}R^{\mu
\nu}\nabla_{\mu}\nabla_{\nu}\\\nonumber
&+&2R\nabla_{\alpha}\nabla_{\beta}-2g_{\alpha\beta}R\nabla^{2}
-4R^{\mu}_{\alpha}\nabla_{\beta}\nabla_{\mu}
-4R^{\mu}_{\beta}\nabla_{\alpha}\nabla_{\mu}-4R_{\alpha\mu \beta
\nu}\nabla^{\mu}\nabla^{\nu})\\\label{2} &\times&f_{G}(G,T),
\end{eqnarray}
where $\nabla^{2}=\Box=\nabla^{l}\nabla_{l}$ and
$G_{\alpha\beta}=R_{\alpha\beta}-\frac{1}{2}Rg_{\alpha\beta}$
demonstrate the d' Alembert operator and the Einstein tensor,
respectively. Moreover,
$\Theta_{\alpha\beta}=-2T_{\alpha\beta}+Pg_{\alpha\beta}$ and
$f(G,T)$ is an arbitrary function of $G$ and $T$ whose partial
derivatives with respect to $G$ and $T$ are expressed by
$f_{G}(G,T)$ and $f_{T}(G,T)$, respectively. The covariant
divergence of Eq.\eqref{2} turns out to be
\begin{eqnarray}\nonumber
\nabla^{\alpha}T_{\alpha\beta}&=&\frac{f_{T}(G,T)}{k^{2}-f_{T}(G,T)}
\left[-\frac{1}{2}g_{\alpha\beta}\nabla^{\alpha}T+(\Theta_{\alpha\beta}+T_{\alpha\beta})\nabla^{\alpha}(\ln
f_{T}(G,T))\right.\\\label{2a}&+&
\left.\nabla^{\alpha}\Theta_{\alpha\beta}\right].
\end{eqnarray}
Equation \eqref{2} can be rewritten in the following form
\begin{equation}\label{3a}
G_{\alpha\beta}=8\pi
T^{(tot)}_{\alpha\beta}=8\pi(T^{(M)}_{\alpha\beta}+T^{(GT)}_{\alpha\beta}),
\end{equation}
where the correction terms of $f(G,T)$ are represented as
\begin{eqnarray}\nonumber
T^{(GT)}_{\alpha\beta}&=&\frac{1}{8\pi}\left[\{(\mu+P)V_{\alpha}V_{\beta}
+\Pi_{\alpha\beta}+q(V_{\alpha}\chi_{\beta}+\chi_{\alpha}V_{\beta})\}f_{T}(G,T)
\right.\\\nonumber
&+&\left.(4R_{\mu\beta}R^{\mu}_{\alpha}+4R^{\mu\nu}R_{\alpha\mu
\beta \nu}-2RR_{\alpha\beta}-2R^{\mu \nu \gamma} _{\alpha}R_{\beta
\mu \nu \gamma})f_{G}(G,T)\right.\\\nonumber
&+&\left.(4g_{\alpha\beta}R^{\mu
\nu}\nabla_{\mu}\nabla_{\nu}-4R_{\alpha\mu
\beta\nu}\nabla^{\mu}\nabla^{\nu}-4R^{\mu}_{\alpha}\nabla_{\beta}\nabla_{\mu}-2g_{\alpha\beta}R\nabla^{2}\right.\\\label{4a}
&+&\left.2R\nabla_{\alpha}\nabla_{\beta}-4R^{\mu}_{\beta}\nabla_{\alpha}\nabla_{\mu}
+4R_{\alpha\beta}\nabla^{2})f_{G}(G,T)\right]+\frac{1}{2}g_{\alpha\beta}f(G,T),
\end{eqnarray}
while the energy-momentum tensor describing the anisotropic and
dissipative matter configuration is given as
\begin{equation}\label{5a}
T^{(M)}_{\alpha\beta} =\mu V_{\alpha}V_{\beta}+Ph_{\alpha\beta}
+\Pi_{\alpha\beta}+q(V_{\alpha}\chi_{\beta}+\chi_{\alpha}V_{\beta}).
\end{equation}
Here, the quantities $V^{\alpha}=\left(A^{-1},0,0,0\right),~
q^{\alpha}=\left(0,qB^{-1},0,0\right)$ and
~$\chi^{\alpha}=\left(0,B^{-1},0,0\right)$ are four velocity, heat
flux and radial four-vector, respectively. Moreover,
\begin{eqnarray}\label{101a}
\Pi_{\alpha\beta}&=&\Pi\left(\chi_{\alpha}\chi_{\beta}-\frac{h_{\alpha\beta}}{3}\right),
\quad  \Pi=P_{r}-P_{\bot},\\\label{101b}
P&=&\frac{P_{r}+2P_{\bot}}{3}, \quad
h_{\alpha\beta}=g_{\alpha\beta}+V_{\alpha}V_{\beta},\\\label{101c}
\chi^{\alpha}\chi_{\alpha}&=&1,\quad \chi^{\alpha}V_{\alpha}=0,\quad
V^{\alpha}q_{\alpha}=0, \quad V^{\alpha}V_{\alpha}=-1.
\end{eqnarray}

The geometry of the non-static dissipative sphere enclosed by a
hypersurface $\Sigma$ is represented by the line element
\begin{equation}\label{8a}
ds^{2}=-A^{2}dt^{2}+B^{2}dr^{2}+C^{2}d\theta^{2}+C^2{\sin^{2}\theta}{d\phi^2},
\end{equation}
where the metric coefficients $A,~B$ and $C$ are functions of $t$
and $r$. The modified field equations in this framework are
evaluated as
\begin{eqnarray}\label{9a}
8\pi(A^{2}\mu+T^{(GT)}_{00}) &=&
\frac{-A^2\left[\frac{2C''}{C}+\frac{C'^2}{C^2}-\frac{B^2}{C^2}-\frac{2C'B'}{CB}\right]}{B^2}
+\frac{\dot{C}(\frac{2\dot{B}}{B}+\frac{\dot{C}}{C})}{C},\\\label{10a}
8\pi (-qAB+T^{(GT)}_{01}) &=&   \frac{2A'\dot C}{AC}+\frac{2C'\dot
B}{CB}-\frac{2\dot C'}{C}, \\\nonumber
8\pi(B^{2}P_{r}+T^{(GT)}_{11})&=&
\frac{-B^2\left[\frac{2\ddot{C}}{C}-\frac{\dot{C}}{C}(\frac{2\dot{A}}{A}
-\frac{\dot{C}}{C})\right]}{A^2}-\frac{B^2}{C^2}+(\frac{C'^2}{C^2}+\frac{2A'C'}{AC}),\\\label{11a}\\\nonumber
8\pi(C^{2}P_\bot+T^{(GT)}_{22})&=&
8\pi\frac{T^{(tot)}_{33}}{\sin^{2}\theta}=\frac{-C^2}{A^2}\left[\frac{\dot{B}\dot{C}}{BC}
+\frac{\ddot B}{B}+\frac{\ddot C}{C}-\frac{\dot A}{A}(\frac{\dot
C}{C}+\frac{\dot B}{B})\right]\\\label{12a}
&+&\frac{C^2}{B^2}\left[(\frac{A'}{A}-\frac{B'}{B})\frac{C'}{C}
-\frac{A'B'}{A B}+\frac{A''}{A}+\frac{C''}{C}\right],
\end{eqnarray}
where dot and prime stand for differentiation with respect to $t$
and $r$, respectively. The terms
$T^{(GT)}_{00},~T^{(GT)}_{01},~T^{(GT)}_{11}$ and $T^{(GT)}_{22}$
include the modified terms whose values are given in
Eqs.(\ref{100})-(\ref{100c}) of Appendix \textbf{A}. The Bianchi
identity has only two non-zero components which are written as
\begin{eqnarray}\nonumber
&&\frac{-1}{A}\left[\dot\mu+A^2{\left({\frac{T^{00(GT)}}{A^4}}\right)^.}
+\left(\mu+\frac{T^{00(GT)}}{A^2}+P_{r}+\frac{T^{11(GT)}}{B^2}\right)\frac{\dot
B}{B}+2\left(\mu\right.\right.\\\nonumber
&&\times\left.\left.\frac{T^{00(GT)}}{A^2}+P_\bot\frac{T^{22(GT)}}{C^2}\right)\frac{\dot
C}{C}\right]-\frac{1}{B}\left(q'-AB\left(\frac{T^{01(GT)}}{A^2B^2}
\right)'+2\left(q\right.\right.\\\label{13a}
&&-\left.\left.\frac{T^{01(GT)}}{AB^2}\right)
\frac{(AC)'}{AC}\right)=Z_{1}+\frac{2\dot
AT^{00(GT)}}{A^4}-\frac{A'T^{01(GT)}}{A^2B^2}-\frac{B'T^{01(GT)}}{AB^3},\\\nonumber
\\\nonumber
&&\frac{1}{A}\left[\dot q-AB\left(\frac{T^{01(GT)}}{A^2B^2}\right)^.
+\left(2q-\frac{2T^{01(GT)}}{AB}\right)\frac{\dot
B}{B}+\left(2q-\frac{2T^{01(GT)}}{AB}\right)\frac{\dot
C}{C}\right]\\\nonumber
&&+\frac{1}{B}\left[P'_{r}+B^2\left(\frac{T^{11(GT)}}{B^4}\right)'
+\left(\mu+\frac{T^{00(GT)}}{A^2}+P_{r}
+\frac{T^{11(GT)}}{B^2}\right)\frac{A'}{A}+\left(2P_{r}\right.\right.
\\\nonumber
&&+\left.\left. \frac{2T^{11(GT)}}{B^2}-2P_{\bot}
-\frac{2T^{22(GT)}}{C^2}\right)\frac{C'}{C}\right]=Z_{2}+\frac{\dot
AT^{01(GT)}}{A^3B}-\frac{2B'T^{11(GT)}}{B^4},\\\label{14a}
\end{eqnarray}
where $Z_{1}$ and $Z_{2}$ consist of the additional curvature terms
given in Eqs.(\ref{100d}) and (\ref{100e}) of Appendix \textbf{A}.
It is noted that the presence of extra force in $f(G,T)$ gravity
results in the non-conservation of the energy-momentum tensor.

The four acceleration, expansion scalar and shear tensor of the
fluid are respectively, given as
\begin{equation}\label{15a}
a_{1}=\frac{A'}{A},\quad
a=\sqrt{a^{\alpha}a_{\alpha}}=\frac{A'}{AB},
\end{equation}
\begin{equation}\label{17a}
\Theta=\left(2\frac{\dot C}{C}+\frac{\dot B}{B}\right)\frac{1}{A},
\end{equation}
\begin{equation}\label{18a}
 \sigma_{11}=\frac{2}{3}B^2\sigma, \quad \sigma_{22}=\frac{\sigma_{33}}{\sin^2
 \theta}=-\frac{1}{3}C^2\sigma,
\end{equation}
\begin{equation}\label{19a}
\sigma^{\alpha\beta}\sigma_{\alpha\beta}=\frac{2}{3}\sigma^2,~\sigma=\left(\frac{\dot
B}{B}-\frac{\dot C}{C}\right)\frac{1}{A}.
\end{equation}
In order to examine the impact of expansion and shear on the matter
distribution, we express Eq.\eqref{10a} as
\begin{equation}\label{21a}
4\pi\left(qB-\frac{T^{(GT)}_{01}}{A}\right)=\frac{1}{3}\left(\Theta-\sigma\right)'-\sigma
\frac{C'}{C}=\frac{C'}{B}\left[\frac{1}{3}D_{C}\left(\Theta-\sigma\right)-\frac{\sigma}{C}\right],
\end{equation}
where $D_{C}=\frac{1}{C'}\frac{\partial}{\partial r}$ is the proper
radial derivative. The mass of a spherical object is defined by
Misner and Sharp \cite{19b} as
\begin{equation}\label{22a}
m=\frac{1}{2}C^3R_{232}^{3}=\left[1-\left(\frac{C'}{B}\right)^2+\left(\frac{\dot
C}{A}\right)^2\right]\frac{C}{2}.
\end{equation}
We introduce the proper time derivative
$(D_{T}=\frac{1}{A}\frac{\partial}{\partial t})$ to analyze the
dynamics of self-gravitating objects. The velocity $(U)$ of the
collapsing matter, defined in terms of areal radius ($C$), is
negative in case of collapse as
\begin{equation}\label{23a}
U=D_{T}C< 0.
\end{equation}
The relation between velocity and mass
of the sphere is given as
\begin{equation}\label{25a}
E\equiv\frac{C'}{B}=\left(1-\frac{2m}{C}+U^2\right)^\frac{1}{2}.
\end{equation}
The proper derivatives (time and radial) of the mass function become
\begin{equation}\label{27a}
D_{T}m=-4\pi\left[\left(P_{r}+\frac{T^{(GT)}_{11}}{B^2}\right)U
+\left(q-\frac{T^{(GT)}_{01}}{AB}\right)E\right]C^2,
\end{equation}
\begin{equation}\label{28a}
D_{C}m=4\pi\left[\left(\mu+\frac{T^{(GT)}_{00}}{A^2}\right)
+\left(q-\frac{T^{(GT)}_{01}}{AB}\right)\frac{U}{E}\right]C^2,
\end{equation}
which lead to
\begin{eqnarray}\nonumber
\frac{3m}{C^3}&=&4\pi\left(\mu+\frac{T^{(GT)}_{00}}{A^2}\right)
-\frac{4\pi}{C^3}\int^{r}_{0}C^3\left[D_{C}\left(\mu+\frac{T^{(GT)}_{00}}{A^2}\right)
\right.\\\label{30a}&-&\left.3\left(q-\frac{T^{(GT)}_{01}}{AB}\right)\frac{U}{CE}\right]C'dr.
\end{eqnarray}

The Weyl tensor $(C^{\lambda}_{\alpha\beta\mu})$ measures the
stretch in a self-gravitating object due to varying gravitational
field of another body. This tensor can be decomposed into magnetic
and electric parts. However, for a spherical system the magnetic
part disappears while electric part is given as
\begin{equation}\label{35a}
E_{\alpha\beta}=C_{\alpha\mu\beta\nu}V^{\mu}V^{\nu}
=\varepsilon(\chi_{\alpha}\chi_{\beta}-\frac{h_{\alpha\beta}}{3}),
\end{equation}
where
\begin{eqnarray}\nonumber
\varepsilon &=&\frac{1}{2A^2}\left[\frac{\ddot C}{C}-\frac{\ddot
B}{B}-\left(\frac{\dot C}{C}+\frac{\dot A}{A}\right)\left(\frac{\dot
C}{C}-\frac{\dot
B}{B}\right)\right]-\frac{1}{2C^2}+\frac{1}{2B^2}\left[\frac{A''}{A}
-\frac{C''}{C}\right.\\\label{36a} &+&\left.\left(\frac{C'}{C}
-\frac{A'}{A}\right)\left(\frac{C'}{C}+\frac{B'}{B}\right)\right].
\end{eqnarray}
The effect of the tidal force on the mass of the fluid distribution
is gauged from the following relation
\begin{equation}\label{46a}
\frac{3m}{C^3}=4\pi\left[\left(\mu+\frac{T^{(GT)}_{00}}{A^2}\right)
-\Pi^{(tot)}\right]-\varepsilon,
\end{equation}
where $\Pi^{(tot)}=\Pi+\Pi^{(GT)}$ and
$\Pi^{(GT)}=(\frac{T^{(GT)}_{11}}{B^2}-\frac{T^{(GT)}_{22}}{C^2})$.
The hypersurface separates the spacetime into two regions, i.e.,
inner and outer regions. The Darmois junction conditions must be
satisfied to prevent any discontinuity at the boundary. The Vaidya
spacetime describes the outer region of a dissipative sphere as
\begin{equation}\label{31a}
ds^{2}=-\left(1-\frac{2M(\upsilon)}{r}\right)dt^{2}
-2drd\upsilon+\left(d\theta^{2}+\sin^{2}\theta
d\phi^{2}\right)r^{2},
\end{equation}
where $M(\upsilon)$ and $\upsilon$ demonstrate the total mass and
retarded time, respectively. The matching of both regions at the
boundary surface is ensured if the following conditions are
fulfilled
\begin{eqnarray}\label{32a}
\left(m\left(t,r\right)\right)_{\Sigma}&=&\left(M\left(\upsilon\right)\right)_{\Sigma},
\\\nonumber
2\left(\frac{\dot C'}{C}-\frac{C'\dot B}{CB}-\frac{\dot C
A'}{CA}\right)_\Sigma&=&\left(-\frac{\left(\frac{2\ddot
C}{C}-\left(\frac{2\dot A}{A}-\frac{\dot C}{C}\right)\frac{\dot
C}{C}\right)B}{A}\right.\\\label{33a}&+&\left.\frac{\left(-\frac{B^2}{C^2}
+\left(\frac{2A'}{A}+\frac{C'}{C}\right)\frac{C'}{C}\right)A}{B}\right)_\Sigma,
\\\label{34a}
\left(q-\frac{T^{(GT)}_{01}}{AB}\right)_{\Sigma}&=&\left(P_{r}
+\frac{T^{(GT)}_{11}}{B^2}\right)_\Sigma,
\end{eqnarray}
where the subscript $\Sigma$ shows that the values are analyzed at
the boundary. In order to discuss the dissipative spherical geometry
along with the conformal flatness condition, Herrera et al.
\cite{25} proposed some analytical solutions to the Einstein
equations. The obtained solutions were matched with the Vaidya
spacetime, and the consequences of relaxational effects on
temperature and evolution of the system have been observed. Tewari
\cite{26} computed the exact solutions of field equations
corresponding to shear-free, dissipative spherical object. At the
boundary, the acquired solutions were compared to the Vaidya metric
and it was observed that after applying some constraints, the
solutions representing the static fluid evolved to radiating
collapse. Vertogradov \cite{27} utilized the general Vaidya metric
to discuss the gravitational collapse by employing an equation of
state and found that it would either be a black hole or a naked
singularity.  

\section{Structure Scalars}

In order to determine the complexity of the system, we split the
Riemann tensor in terms of structure scalars by following Herrera's
technique \cite{21}. The disintegrated form of the Riemann tensor in
terms of Ricci scalar, Ricci and Weyl tensors reads
\begin{equation}\label{90a}
R^{\rho}_{\alpha\beta\mu}=C^{\rho}_{\alpha\beta\mu}
+\frac{1}{2}R_{\alpha\mu}\delta^{\rho}_{\beta}
+\frac{1}{2}R^{\rho}_{\beta}g_{\alpha\mu}
-\frac{1}{2}R_{\alpha\beta}\delta^{\rho}_{\mu}-\frac{1}{2}R^{\rho}_{\mu}g_{\alpha\beta}
-\frac{1}{6}R\left(\delta^{\rho}_{\beta}g_{\alpha\mu}\right.
-\left.g_{\alpha\beta}\delta^{\rho}_{\mu}\right),
\end{equation}
which is rewritten as
\begin{equation}\label{39a}
R^{\alpha\gamma}_{\beta\delta}=C^{\alpha\gamma}_{\beta\delta}+16\pi
T^{(tot)[\alpha}_{[\beta}\delta^{\gamma]}_{\delta]}+8\pi
T^{(tot)}\left(\frac{1}{3}\delta^{\alpha}_{[\beta}\delta^{\gamma}_{\delta]}-\delta^{[\alpha}_{[\beta}\delta^{\gamma]}_{\delta]}\right).
\end{equation}
We introduce the tensors $Y_{\alpha\beta}$ and $X_{\alpha\beta}$,
respectively defined as
\begin{eqnarray}\label{37a}
Y_{\alpha\beta} &=&R_{\alpha\gamma\beta\delta}V^{\gamma}V^{\delta},
\\\label{38a} X_{\alpha\beta}
&=&^{\ast}R^{\ast}_{\alpha\gamma\beta\delta}V^{\gamma}V^{\delta}=\frac{1}{2}\eta^{\epsilon\mu}_{\alpha\gamma}
R^{\ast}_{\epsilon\mu\beta\delta}V^{\gamma}V^{\delta},
\end{eqnarray}
where $\eta^{\epsilon\mu}_{\alpha\gamma}$ denotes the Levi-Civita
symbol and $R^{\ast}_{\alpha\beta\gamma\delta}
=\frac{1}{2}\eta_{\epsilon\mu\gamma\delta}R^{\epsilon\mu}_{\alpha\beta}$.
In electrodynamics, the tensor $Y_{\alpha\beta}$ specifies the
electric component of the Riemann tensor whereas the tensor
$X_{\alpha\beta}$ has no analogy. Dual provides a simple
constitutive relation for free space, with vacuum having its own
meaningful electric and magnetic fields. The tensors
$Y_{\alpha\beta}$ and $X_{\alpha\beta}$ will later be expressed in
terms of matter variables such as energy density $\mu$, anisotropic
pressure $\Pi$ and heat flux $q$ as these tensors are defined in the
form of Riemann tensor. These tensors are written in the combination
of their trace-free $(Y_{TF}, X_{TF})$ and trace parts $(Y_{T},
X_{T})$   as
\begin{eqnarray}\label{40a}
Y_{\alpha\beta}
&=&\frac{h_{\alpha\beta}Y_{T}}{3}+(\chi_{\alpha}\chi_{\beta}-\frac{h_{\alpha\beta}}{3})Y_{TF},
\\\label{41a}
X_{\alpha\beta}
&=&\frac{h_{\alpha\beta}X_{T}}{3}+(\chi_{\alpha}\chi_{\beta}-\frac{h_{\alpha\beta}}{3})X_{TF}.
\end{eqnarray}
The structure scalars corresponding to the current setup are
obtained as
\begin{eqnarray}\label{42a}
Y_{T}&=& 4\pi\left(\mu+3P_{r}-2\Pi
\right)+\frac{\left(\mu+P\right)f_{T}}{2}+M^{(GT)},  \\\label{43a}
Y_{TF} &=& \varepsilon-4\pi
\Pi+\frac{\Pi}{2}f_{T}+L^{(GT)},\\\label{44a} X_{T}&=& 8\pi
\mu+Q^{(GT)},  \\\label{45a} X_{TF} &=&-\varepsilon-4\pi
\Pi+\frac{\Pi}{2}f_{T},
\end{eqnarray}
where
$L^{(GT)}=\frac{J^{(GT)}_{\alpha\beta}}{s_{\alpha}s_{\beta}-\frac{1}{3}h_{\alpha\beta}}$.
The terms $M^{(GT)}$, $J^{(GT)}_{\alpha\beta}$ and $Q^{(GT)}$ are
provided in Appendix \textbf{B}. The scalars $X_{T}$ and $Y_{T}$ are
responsible for the energy density and local anisotropic pressure,
respectively. Employing Eqs.\eqref{36a} and \eqref{43a}, the scalar
function $Y_{TF}$ is expressed as
\begin{eqnarray}\nonumber
Y_{TF}&&=-4\pi\Pi-4\pi\Pi^{(GT)}-L^{(GT)}+\frac{4\pi}{C^3}
\int C^3\left[D_{C}\left(\mu+\frac{T^{(GT)}_{00}}{A^2}\right)\right.
\\\label{47a}&&\left.-3\left(q-\frac{T^{(GT)}_{01}}{AB}\right)\frac{U}{CE}\right]C'dr,
\end{eqnarray}
where $\Pi,\mu$ and $q$ represent the anisotropic pressure, energy
density and heat flux, respectively. The above equation shows that
$Y_{TF}$ involves the contribution from inhomogeneous energy
density, heat flux and anisotropic pressure. The scalar $X_{TF}$
measures the inhomogeneity within the system in $f(G,T)$ gravity  as
\begin{equation}\label{48a}
X_{TF}=4\pi\Pi^{(GT)}-\frac{4\pi}{C^3}\int
C^3\left[D_{C}\left(\mu+\frac{T^{(GT)}_{00}}{A^2}\right)
-3\left(q-\frac{T^{(GT)}_{01}}{AB}\right)\right.
\left.\frac{U}{CE}\right]C'dr.
\end{equation}

\section{Evolution Modes}

Different interrelated physical factors such as pressure and energy
density play an important role in the complex nature of the cosmic
structure. The scalar $Y_{TF}$ involves energy density
inhomogeneity, heat flux and pressure anisotropy along with $f(G,T)$
corrections. Therefore, we identify $Y_{TF}$ as the complexity
factor for the fluid distribution in a non-static system. The
condition $Y_{TF}=0$ corresponds to complexity free system. As fluid
evolves with time, we need to determine its pattern of evolution.
For this purpose, we discuss two modes of evolution, i.e.,
homologous evolution and homogeneous expansion in the subsequent
sections and choose the simplest mode to minimize complexity.

\subsection{Homologous Evolution}

The word homologous means self-similar or having the same pattern.
Therefore, homologous collapse refers to a linear relation between
velocity and the radial distance, i.e.,  matter is pulled to the
core at the same rate throughout the collapse of astrophysical
objects. The object undergoing homologous collapse emits strong
gravitational radiation as compared to the object whose core
collapses first. We rewrite Eq.\eqref{21a} as
\begin{equation}\label{52a}
D_{C}\left(\frac{U}{C}\right)=\frac{4\pi}{E}\left(q-\frac{T^{(GT)}_{01}}{AB}\right)
+\frac{\sigma}{C},
\end{equation}
which after integration produces
\begin{equation}\label{53a}
U=h(t)C+C\int^{r}_{0}\left[\frac{4\pi}{E}\left(q-\frac{T^{(GT)}_{01}}{AB}\right)
+\frac{\sigma}{C}\right]C'dr,
\end{equation}
where $h(t)$ is an integration constant whose evaluation at the
boundary yields
\begin{equation}\label{54a}
U=\frac{U_{\Sigma}}{C_{\Sigma}}C-C\int^{r_{\Sigma}}_{r}
\left[\frac{4\pi}{E}\left(q-\frac{T^{(GT)}_{01}}{AB}\right)
+\frac{\sigma}{C}\right]C'dr.
\end{equation}
Heat dissipation and shear scalar cause the fluid to deviate from
homologous mode. Therefore, if the terms in the integral cancel each
other then $U\sim C$ which is the required condition for homologous
evolution \cite{22a} which provides $U=h(t)C$ with
$h(t)=\frac{U_{\Sigma}}{C_{\Sigma}}$. The homologous condition in
$f(G,T)$ gravity is evaluated as
\begin{equation}\label{56a}
\frac{4\pi
B}{C'}\left(q-\frac{T^{(GT)}_{01}}{AB}\right)+\frac{\sigma}{C}=0.
\end{equation}

\subsection{Homogeneous Expansion}

The condition for homogeneous expansion reads $\Theta'=0$. In other
words, the rate of collapse or expansion of the cosmic objects is
homogeneous if it does not depend upon $r$. Applying this condition
on Eq.\eqref{21a} leads to
\begin{equation}\label{57a}
4\pi\left(q-\frac{T^{(GT)}_{01}}{AB}\right)=-\frac{C'}{B}
\left[\frac{1}{3}D_{C}(\sigma)+\frac{\sigma}{C}\right].
\end{equation}
Comparing the above equation with \eqref{56a} yields
\begin{equation}\label{58a}
D_{C}(\sigma)=0.
\end{equation}
Moreover, the regularity conditions at the middle yield $\sigma=0$.
Thus, Eq.(\ref{57a}) takes the form
\begin{equation}\label{59a}
q=\frac{T^{(GT)}_{01}}{AB},
\end{equation}
which implies that the fluid is dissipative. However, in the context
of GR, the homogeneous pattern of evolution yields a shear-free and
non-dissipative fluid distribution.

\section{Kinematical and Dynamical Variables}

We select the simplest mode of evolution by investigating the
behavior of different physical variables. For sake of simplifying
the calculations, we assume $C$ as a separable function of $r$ and
$t$. The homologous condition along with Eq.\eqref{21a} leads to
\begin{equation}\label{62a}
\left(\Theta-\sigma\right)'=\left(\frac{3\dot C}{AC}\right)'=0.
\end{equation}
This implies that the fluid is geodesic as $A'=0 ~(a=0)$. We take
$A=1$ without any loss of generality. Conversely, the geodesic
condition provides
\begin{equation}\label{63a}
\Theta-\sigma=\frac{3\dot C}{C}.
\end{equation}
The successive derivatives of the above equation closer to the core
lead to the homologous condition. Thus, the geodesic condition
implies a homologous fluid and vice-versa. In GR, the fluid
distribution is shear-free when $q=0$. However, in $f(G,T)$ theory
the shear scalar for $q=0$ is evaluated as
\begin{equation}\label{64a}
\sigma=4\pi\frac{T^{(GT)}_{01}C}{C'}.
\end{equation}
If a non-dissipative fluid collapses homogeneously, we obtain
$T^{(GT)}_{01}=0$ while Eq.\eqref{57a} yields the shear scalar as
\begin{equation}\label{65a}
\sigma=\frac{12 \pi}{C^3}\int \frac{C^3T^{(GT)}_{01}}{A}dr+\frac{b
(t)}{C^3},
\end{equation}
where arbitrary integration function ($b(t)$) must be zero at the
center $(C=0)$. Thus, homogeneous expansion infers homologous
evolution for the non-dissipative case as  $T^{(GT)}_{01}=0
\Rightarrow \sigma=0\Rightarrow U\sim C$. As homogeneous expansion
implies homologous evolution, this shows that the homologous
evolution is the simplest mode. When a fluid undergoes homologous
evolution, the relation between mass and velocity during collapse
becomes
\begin{equation}\label{67a}
D_{T}U=-\frac{m}{C^2}-4\pi\left(P_{r}+\frac{T^{(GT)}_{11}}{B^2}\right)C.
\end{equation}
Using the definition of $Y_{TF}$ in the above expression yields
\begin{equation}\label{68a}
\frac{3D_{T}U}{C}=Y_{TF}-L^{(GT)}+4\pi\Pi^{(GT)}-4\pi
\left[\left(\mu+\frac{T^{(GT)}_{00}}{A^2}\right)+3\left(P_{r}\right.\right.
+\left.\left.\frac{T^{(GT)}_{11}}{B^2}\right)-2\Pi\right],
\end{equation}
which leads to
\begin{equation}\label{71a}
\frac{\ddot C}{C}-\frac{\ddot B}{B}=Y_{TF}-L^{(GT)}+4\pi\Pi^{(GT)}.
\end{equation}
Thus, the matter distribution is free from complexity when
$\frac{\ddot C}{C}-\frac{\ddot B}{B}+L^{(GT)}-4\pi\Pi^{(GT)}=0$.
Integration of Eq.(\ref{71a}) for the case $Y_{TF}=0$ produces
\begin{eqnarray}\label{73a}
B=C_{1}(t)\left[g_{1}(r)\int \frac{1}{C_{1}(t)^2}
e^{\int \left(L^{(GT)}-4\pi\Pi^{(GT)}\right)\frac{C_{1}(t)}{\dot C_{1}(t)}dt}
dt+g_{2}(r)\right],
\end{eqnarray}
where $g_{1}(r)$ and $g_{2}(r)$ are integration functions. The above
equation can be conveniently written as
\begin{eqnarray}\nonumber
B&=&C_{1}(t)C'_{2}(r)\left[\tilde g_{1}(r)\int
\frac{1}{C_{1}(t)^2}e^{\int
\left(L^{(GT)}-4\pi\Pi^{(GT)}\right)\frac{C_{1}(t)}{\dot
C_{1}(t)}dt}dt +\tilde g_{2}(r)\right]\\\label{}&=&ZC',
\end{eqnarray}
where $Z=\tilde g_{1}(r)\int \frac{1}{C_{1}(t)^2}e^{\int
\left(L^{(GT)}-4\pi\Pi^{(GT)}\right)\frac{C_{1}(t)}{\dot
C_{1}(t)}dt}dt+\tilde g_{2}(r)$, $\tilde g_{1}(r)$=
$\frac{g_{1}(r)}{C'_{2}(r)}$ and $\tilde g_{2}(r)$=
$\frac{g_{2}(r)}{C'_{2}(r)}$.

We now propose possible solutions for non-dissipative as well as
dissipative fluids satisfying the homologous and vanishing
complexity conditions. As the considered system is non-static and
$f(G,T)$ field equations are highly non-linear, it is convenient to
choose a linear $f(G,T)$ model of the form $f(G,T)=h_1(G)+h_2(T)$,
where $h_1(G)=\xi G^n$ and $h_2(T)=\kappa T$ \cite{23}. Further,
$\xi$ and $\kappa$ are real numbers while $n>0$. For the present
work, we select $\xi=n=1$. One of the extension of modified
Gauss-Bonnet gravity is $f(G,T)$ gravity. The term GB alone is a
total derivative and can be integrated out leaving no modified
results. However, in $f(G,T)$ theory, it encompasses the effects of
both GB invariant and $T$ which does not correspond to the total
derivative and therefore cannot be integrated out in the action as a
surface term. The model mentioned in the manuscript is not the same
as GR because it includes the contribution from the GB term as well
as $T$. Hence, the model $\xi G+\kappa T$ cannot be considered as GR
version. In order to inspect the complexity of self-gravitating
objects, some other $f(G,T)$ models \cite{28} can also be utilized
such as
\begin{itemize}
\item $f(G,T)=G^2+2T$
\item $f(G,T)=f(G)+\lambda T$.
\end{itemize}
At the current stage, the system contains two unknowns: $B(t,r)$ and
$C(t,r)$ which can be determined via the homologous and vanishing
complexity conditions. For the case of non-dissipative fluid, these
conditions are computed as
\begin{eqnarray}\nonumber
&&\frac{1}{{(\kappa +8 \pi ) B^3 C^2}}\left[B\left(\dot B C'\ddot
C-2 C \left(\dot B C'+5 \ddot B \dot C'\right)\right)+10 \dot B
\ddot B C C'\right.\\\label{77a}&&\left.+2 B^2 \dot C' \left(C-4
\ddot C\right)\right]=0,\\\nonumber &&\frac{1}{B C}\left[-2 \dot B
B^3 \dot C \left(C-4 \ddot C\right)-C'B \left(-32 \dot B \dot
C'^2+\left(3 \ddot B+\dot B^2\right)
C'\right.\right.\\\nonumber&&\left.\left.+2 B' C\right)+4 C'^2\dot
B^2-2 B^2 \left(-C C''-2 \dot C+8 \dot
C'^2+C'^2\right)\right.\\\label{78a}&&\left.+2 B^4 \left(\dot
C^2+1\right)\right]=0.
\end{eqnarray}
In dissipative case, the obtained vanishing complexity condition is
the same as \eqref{78a} while the homologous condition is given as
\begin{eqnarray}\nonumber
&&q=\frac{1}{16\pi^2(\kappa +1) B^5 C^4}\left[C'\left(\dot BC-B \dot
C\right) \left(B \left(C \left(\dot B C'+40 \pi\dot C' \ddot B
\right)\right.\right.\right.\\\label{79a}&&\left.\left.\left.-4
\pi\ddot C \dot B C'\right)-40 \pi \dot B \ddot B C C'+B^2
\left(-\dot C'\right) \left(C-32 \pi \ddot C\right)\right)\right].
\end{eqnarray}
The homologous condition and vanishing complexity condition for the
non-dissipative and dissipative cases in the context of GR are as
follows
\begin{equation}\label{102a}
\frac{1}{4 \pi B^2 C}\left(B\dot C'-\dot B C'\right)=0,
\end{equation}
\begin{equation}\label{102b}
\frac{B^2 \left(\dot C^2+1\right)}{C}+C''=\frac{B' C'}{B}+B \dot B
\dot C+\frac{C'^2}{C},
\end{equation}
\begin{equation}\label{102c}
q=\frac{1}{16 \pi ^2 B^4 C^3}\left(C' \left(B \dot C-\dot B C\right)
\left(B \dot C'-\dot B C'\right)\right).
\end{equation}
\begin{figure}\center
\epsfig{file=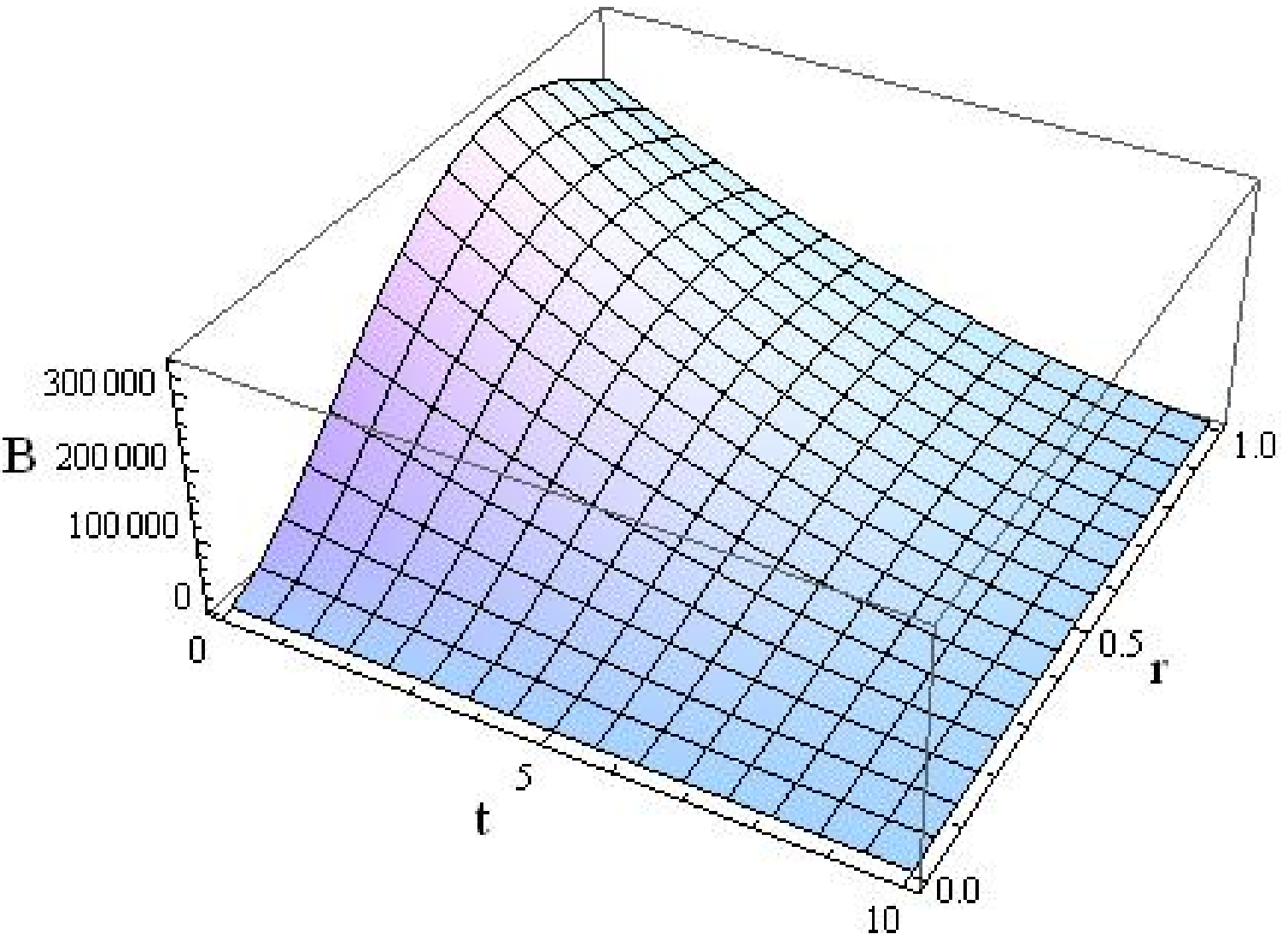,width=0.4\linewidth}\epsfig{file=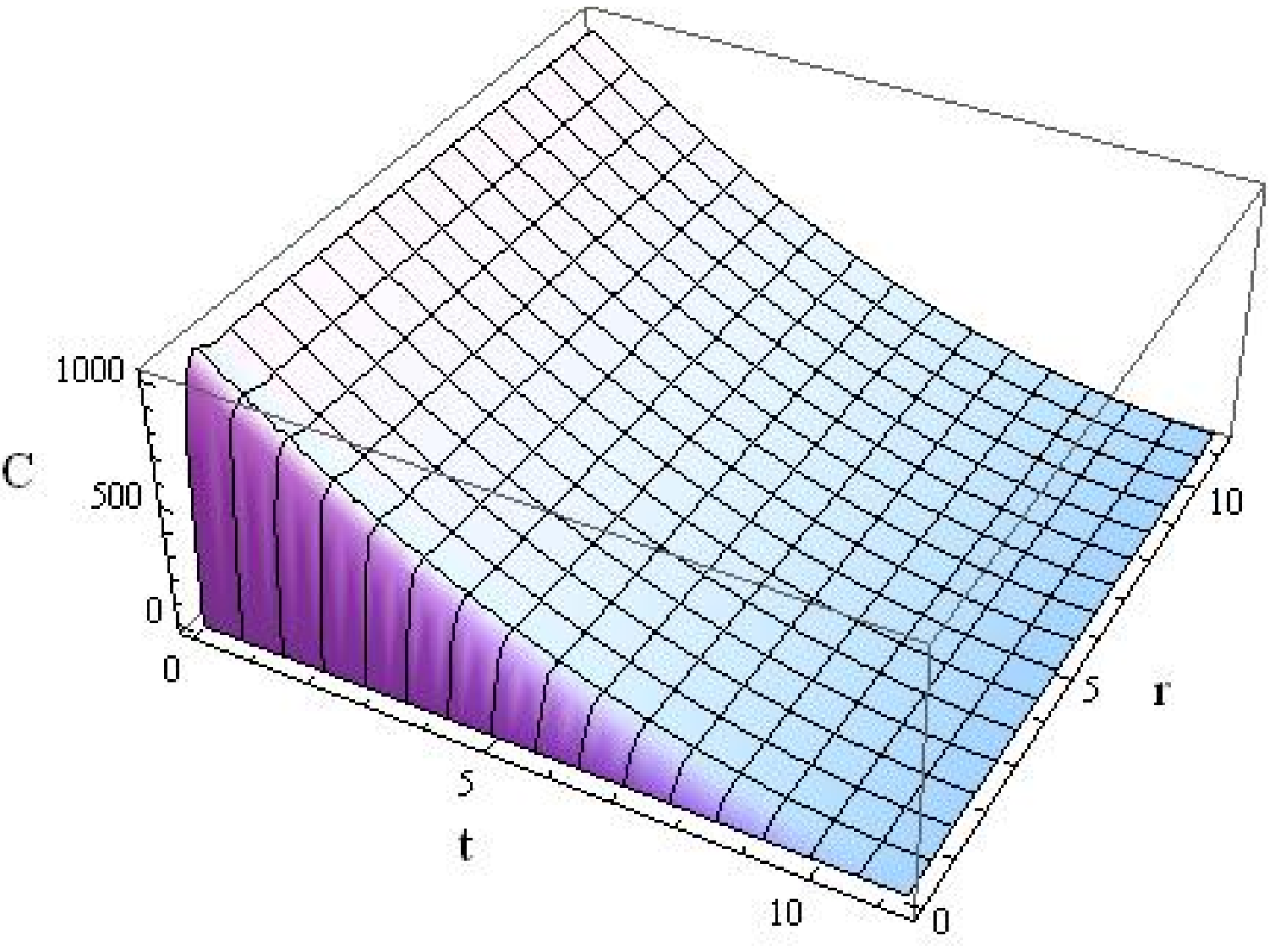,width=0.4\linewidth}
\epsfig{file=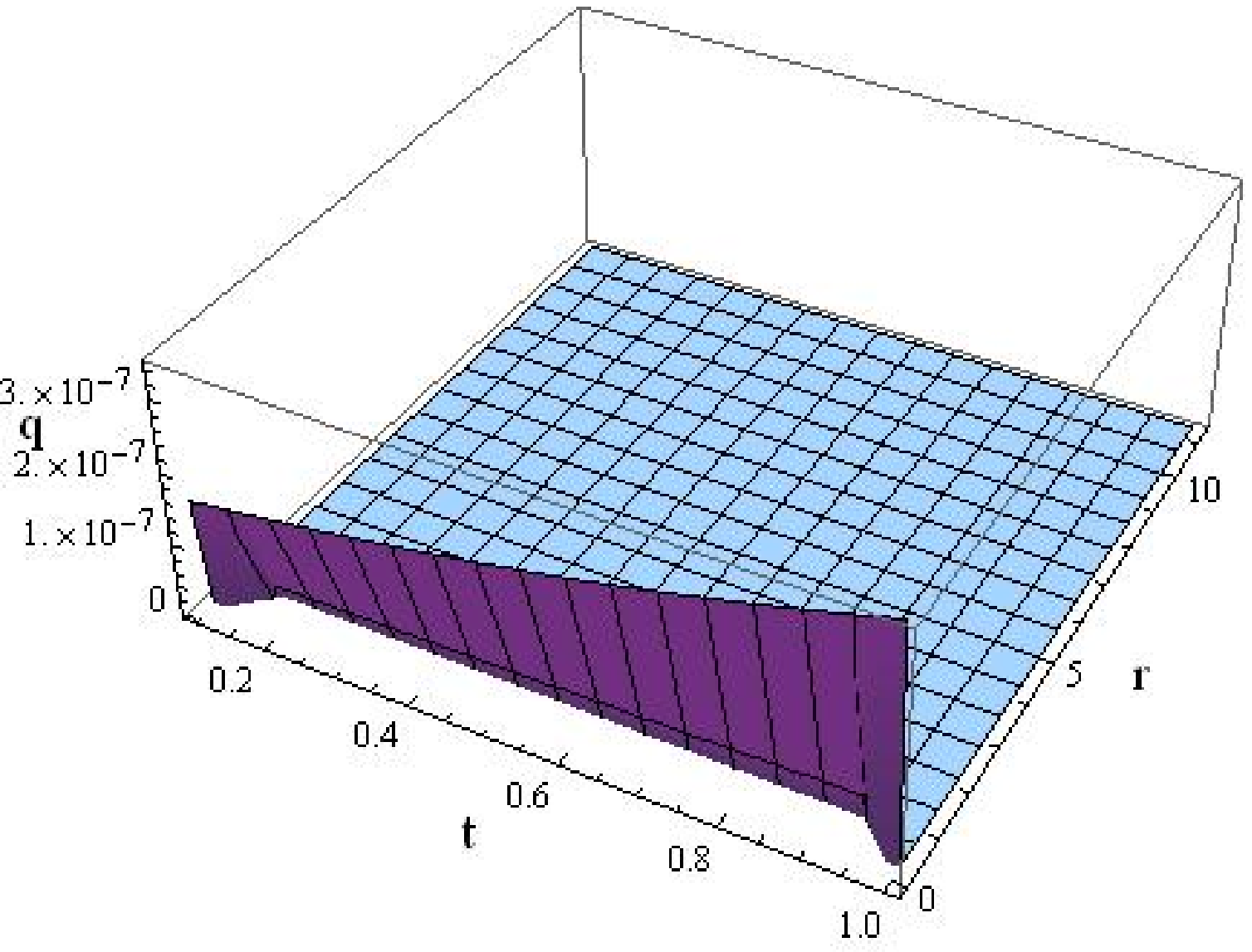,width=0.4\linewidth} \caption{Plots of metric
potentials and heat flux versus $r$ and $t$ in $f(G,T)$ framework.}
\end{figure}

In order to compare the complex nature of self-gravitating fluids in
$f(G,T)$ gravity, we have shown the behavior of $B,C$ and $q$ in the
non-static scenario in Figure \textbf{1}. In the upper left graph,
the value of $B$ is slightly reduced by increasing the value of $t$
while the increment in $r$ makes B maximum. The graph of $C$ (upper
right) indicates that its value gradually decreases with increasing
$t$ and remains at its maximum throughout the $r$ range. The
graphical analysis of $q$ illustrates that it has more dissipation
at the center as compared to GR with respect to $t$ and it goes on
increasing at the boundary. Radially, it is maximum at the center
and then begins to decrease for larger values of $r$. Figure
\textbf{2} exhibits the functioning of $B,C$ and $q$ in the context
of GR. The metric coefficient $B$ linearly increases with $t$
whereas its value monotonically increases towards the boundary with
$r$. The graphic representation of $C$ depicts the similar behaviour
as in $f(G,T)$ with respect to $t$ whereas it first increases and
then shifts to the decreasing behavior with $r$. The lower graph in
Figure \textbf{2} interprets that heat flux is minimum at the center
and becomes maximum at the larger values of $t$, while it shows the
similar pattern as in $f(G,T)$ for the radial trend.

\section{Stability of Zero Complexity Condition}

Following the procedure in \cite{21}, the evolution equation is
attained in terms of complexity factor by using Eqs.\eqref{13a},
\eqref{43a} and \eqref{45a}
\begin{figure}\center
\epsfig{file=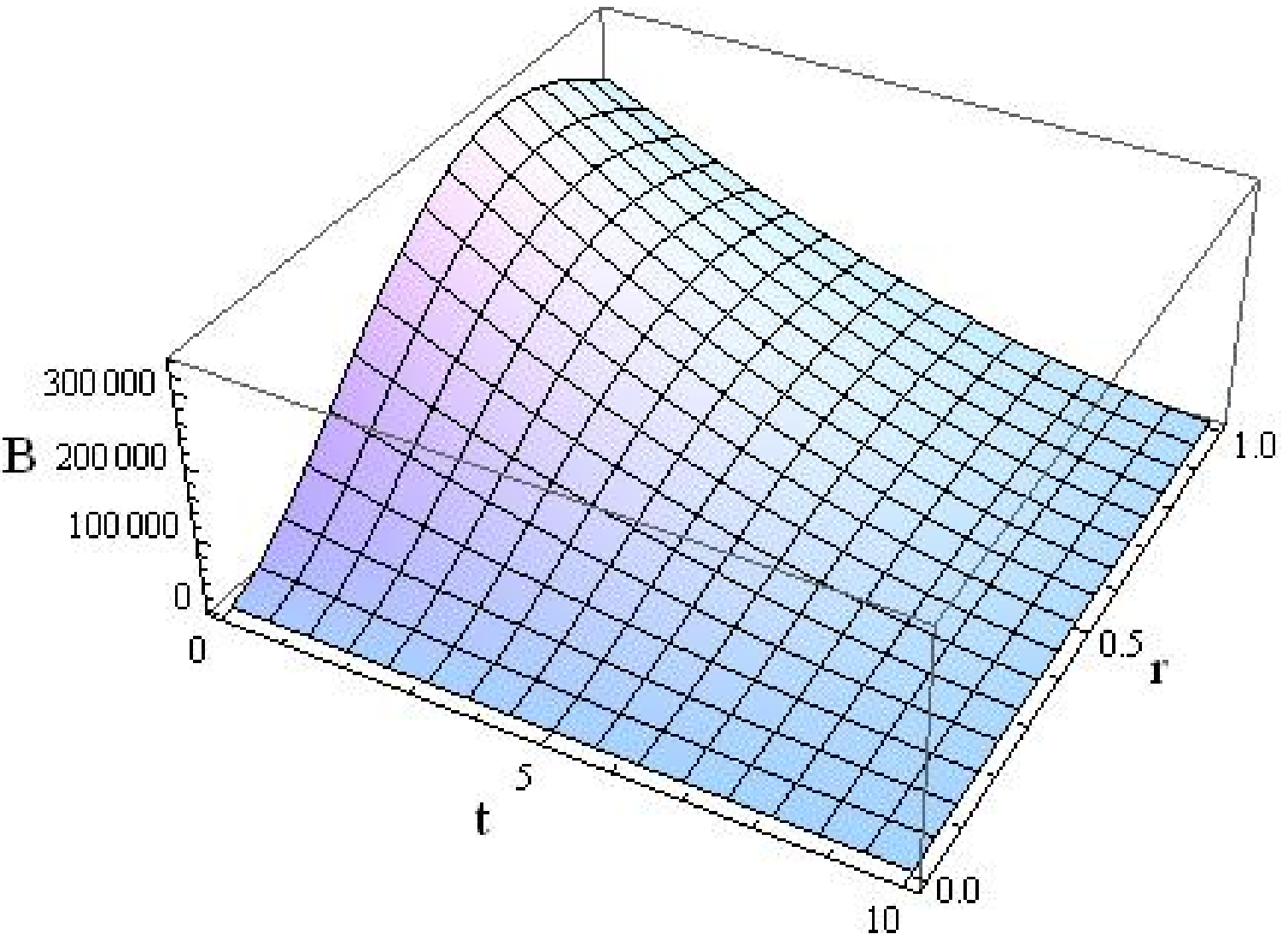,width=0.4\linewidth}\epsfig{file=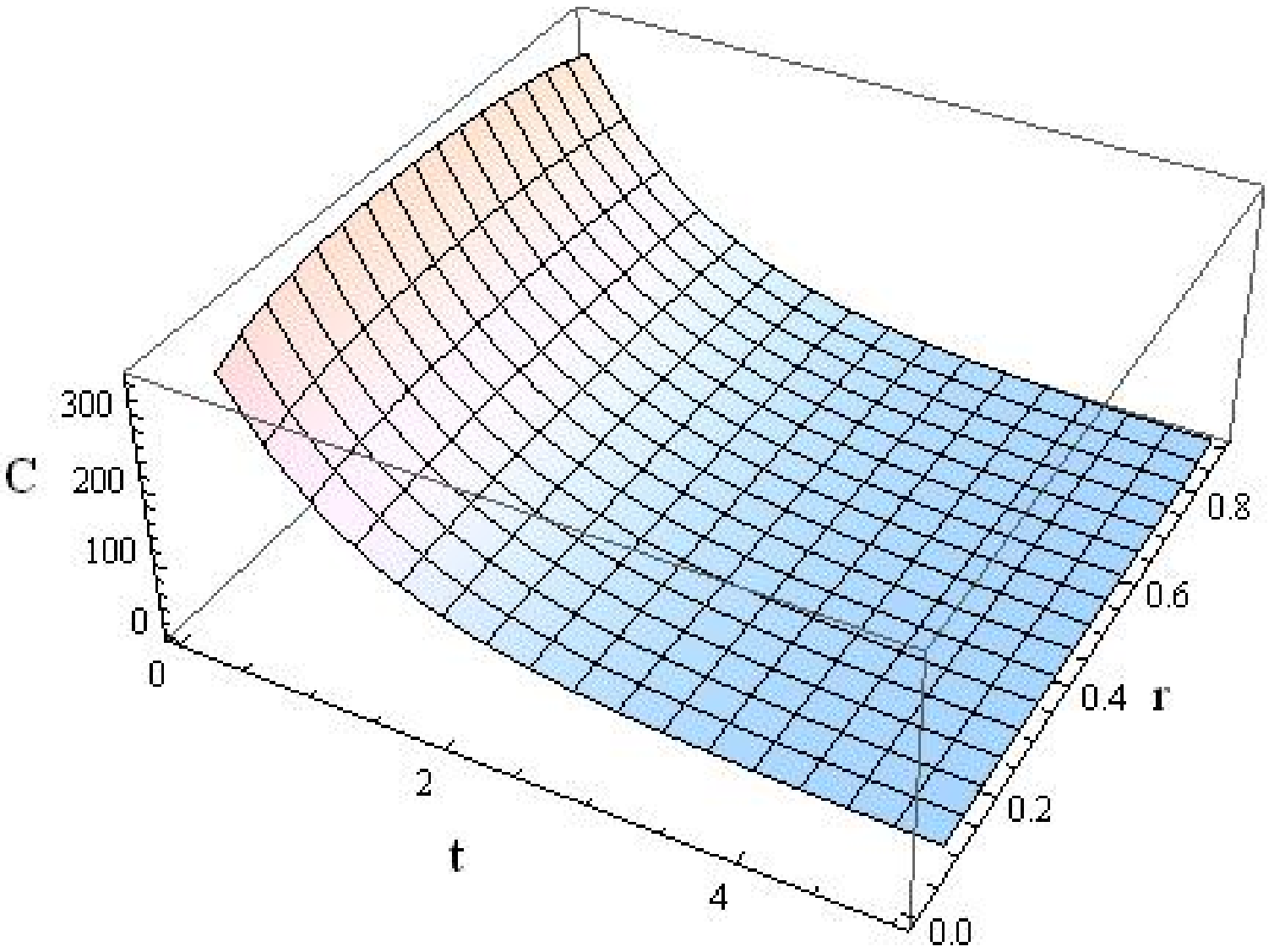,width=0.4\linewidth}
\epsfig{file=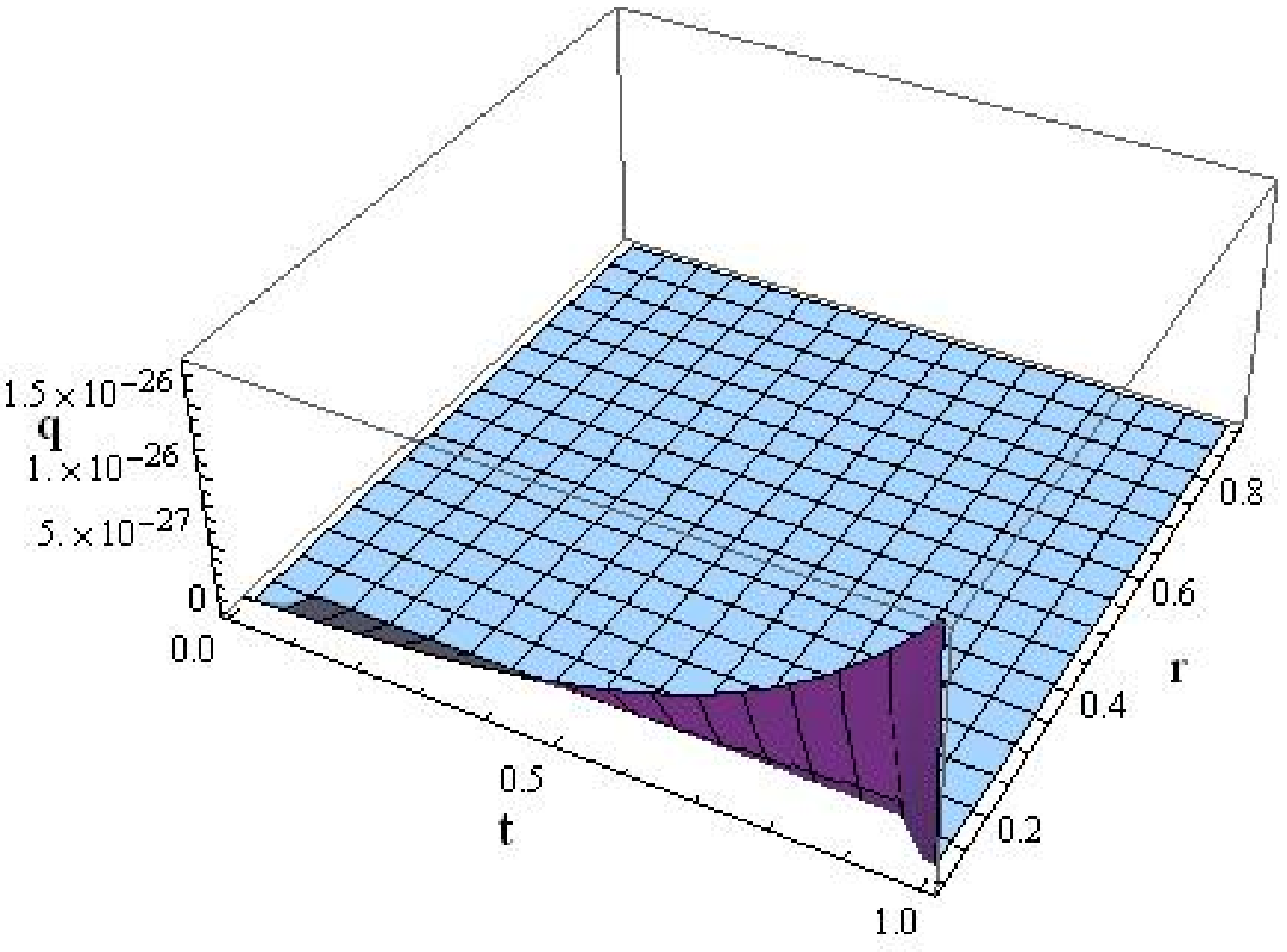,width=0.4\linewidth} \caption{Plots of metric
potentials and heat flux versus $r$ and $t$ in GR framework.}
\end{figure}
\begin{eqnarray}\nonumber
&&-4\pi\left(\mu+T^{00(GT)}+P_{r}+\frac{T^{11(GT)}}{B^2}\right)\sigma
-\frac{4\pi}{B}\left[q'-B(\frac{T^{00(GT)}}{B^2})'
\right.\\\nonumber&&\left.-(q-\frac{T^{01(GT)}}{B^2})\frac{C'}{C}\right]
-\dot Y_{TF}-\dot L^{(GT)}-8\pi\dot \Pi-4\pi\dot\Pi^{(GT)}-4\pi
\left(T^{(GT)}_{00}\right)^.\\\nonumber&&-3\frac{\dot
C}{C}\left(Y_{TF}-L^{(GT)}\right)-4\pi Z_{1}+\frac{4\pi
T^{00(GT)}B'}{B^2}-\frac{12\pi T^{00(GT)}\dot
C}{C}\\\label{87a}&&+\left(12\pi
T^{(GT)}_{00}-12\pi\Pi^{(GT)}+\frac{12\pi
T^{(GT)}_{11}}{B^2}\right)\frac{\dot
C}{C}-\frac{16\pi\Pi^{(tot)}\dot C}{C} =0.
\end{eqnarray}
At the initial time $t=0$, the substitution
$Y_{TF}=\Pi^{(tot)}=q=\sigma=0$ for the non-dissipative case yields
\begin{eqnarray*}\nonumber
&&-\dot Y_{TF}+4\pi(\frac{T^{00(GT)}}{B^2})'-4\pi
\left(T^{(GT)}_{00}\right)^.-\dot L^{(GT)}-8\pi\dot
\Pi-4\pi\dot\Pi^{(GT)}+3L^{(GT)}\\\nonumber&&\times\frac{\dot
C}{C}-4\pi Z_{1}+\frac{4\pi T^{00(GT)}B'}{B^2}-\frac{12\pi
T^{00(GT)}\dot C}{C}+\left(12\pi T^{(GT)}_{00}+\frac{12\pi
T^{(GT)}_{11}}{B^2}\right)\\&&\times\frac{\dot C}{C}=0.
\end{eqnarray*}
Inserting the above condition in the derivative of Eq.\eqref{47a} at
$t=0$ leads to
\begin{eqnarray}\nonumber
&&\frac{4\pi}{C^3}\int
^{r}_{0}[(\mu+\frac{T^{(GT)}_{00}}{A^2})^.]'=+4\pi(\frac{T^{00(GT)}}{B^2})'-4\pi
\left(T^{(GT)}_{00}\right)^.-\frac{12\pi T^{00(GT)}\dot
C}{C}\\\label{90b}&&+\frac{4\pi T^{00(GT)}B'}{B^2}-4\pi
Z_{1}+3L^{(GT)}\frac{\dot C}{C}+(12\pi T^{(GT)}_{00}+\frac{12\pi
T^{(GT)}_{11}}{B^2})\frac{\dot C}{C}.
\end{eqnarray}
Moreover, differentiation of Eq.\eqref{87a} with respect to time
along with the condition $Y_{TF}=\sigma=0$ is expressed as
\begin{eqnarray}\nonumber
&&-\ddot
Y_{TF}+\left[4\pi\left(\frac{T^{00(GT)}}{B^2}\right)'\right]^.-4\pi\left(T^{(GT)}_{00}\right)^{..}-\ddot
L^{(GT)}-4\pi \dot Z_{1}+\dot
L^{(GT)}\\\nonumber&&-4\pi\ddot\Pi^{(GT)}-12\pi\left(\frac{T^{00(GT)}\dot
C}{C}\right)^.+\frac{3\dot
C}{C}[-4\pi\left(\frac{T^{00(GT)}}{B^2}\right)'+4\pi\left(T^{(GT)}_{00}\right)^{.}\\\nonumber&&+4\pi
Z_{1}-4\pi\left(\frac{B'T^{00(GT)}}{B^2}\right)+4\pi\dot\Pi^{(GT)}+12\pi\left(\frac{T^{00(GT)}\dot
C}{C}\right)-3L^{(GT)}\frac{\dot C}{C}\\\nonumber&&-\left(12\pi
T^{(GT)}_{00}+\frac{12\pi T^{(GT)}_{11}}{B^2}\right)\frac{\dot
C}{C}+8\pi\dot\Pi]+3\left(\frac{L^{(GT)}\dot
C}{C}\right)^.-\frac{16\pi\Pi^{(tot)}\dot
C}{C}\\\label{89a}&&-8\pi\ddot\Pi+\left[(12\pi
T^{(GT)}_{00}+\frac{12\pi T^{(GT)}_{11}}{B^2})\frac{\dot
C}{C}\right]^.+4\pi\left(\frac{B'T^{00(GT)}}{B^2}\right)^.=0.
\end{eqnarray}

The higher derivatives of Eq.\eqref{47a} along with (\ref{89a}) show
that the inhomogeneous energy density, anisotropic pressure and dark
source terms lead to the instability of vanishing complexity
condition. The state with zero complexity is achieved and maintained
in the scenario of isotropic pressure, homogeneous energy density
with no modified terms. Furthermore, the presence of heat flux also
influences the stability of the zero complexity factor in the
dissipative mode. In the present manuscript, the vanishing
complexity factor condition $(Y_{TF}=L^{(GT)}-4\pi\Pi^{(GT)})$ is
determined which comes out in the form of matter variables and extra
curvature terms. With the help of experimental data of any compact
objects regarding mass and radius, this condition can be found
helpful to check the stability and viability of that particular
object. Different techniques have been employed to analyze the
stability and viability of various astrophysical objects \cite{24}.

\section{Conclusions}

The focus of this paper is to determine the complexity of a
non-static sphere in the background of $f(G,T)$ gravity. For this
purpose, we have assumed an inhomogeneous sphere with heat flux and
anisotropic pressure. We have adopted Herrera's technique to split
the Riemann tensor into four scalars which specify the structure of
self-gravitating objects. We have chosen $Y_{TF}$ as the complexity
factor based on the following reasons.
\begin{enumerate}
\item In the static case, it has already served as an appropriate
measure of complexity \cite{12}.
\item It is the only factor which incorporates the energy density
inhomogeneity, anisotropic pressure, heat dissipation and correction
terms.
\end{enumerate}
We have inspected two possible modes of evolution namely homologous
and homogeneous patterns. Choosing homologous mode as the simplest
pattern of evolution, we have derived the solutions for dissipative
and non-dissipative cases by applying the condition of zero
complexity. We have interpreted the numerical solutions through
graphical analysis in GR as well as $f(G,T)$ scenarios to
distinguish the behavior of complexity between them. We have also
discussed the factors which cause the system to deviate from
vanishing complexity condition during the evolution process.

The structure scalars in $f(G,T)$ theory include the contribution
from dark source terms and potential functions. Consequently, the
presence of higher order curvature terms produce complexity in the
dynamical sphere. Moreover, the geodesic nature of the fluid
evolving homologously (in dissipative and non-dissipative scenarios)
leads to the choice of homologous pattern as the simplest mode of
evolution. We have devised the homologous condition in this theory
which includes the modified terms. Further, the condition $Y_{TF}=0$
was achieved in GR when $\mu=\Pi=q=0$ whereas, in $f(G,T)$ theory,
an extra constraint $L^{(GT)}-4\pi\Pi^{(GT)}=0$ along with the
aforementioned conditions provide a complexity free system.
Moreover, the shear tensor does not vanish for the non-dissipative
case in contrast to GR. The modified homologous and vanishing
complexity conditions have been used to evaluate the possible
solutions for non-dissipative and dissipative models. Isotropic
pressure and homogeneous energy density lead to the stability of
vanishing complexity condition in the absence of correction terms
and heat flux. We conclude from Eq.(\ref{43a}) that the $f(G,T)$
system is more complex as compared to its GR analog. In the
formalism of $f(G,T)$ gravity, the complexity of self-gravitating
system for the static case has been evaluated \cite{17b} and the
core equation (vanishing complexity condition) of the complexity
work is recovered from the non-static system. It is worthwhile to
mention here that our all results reduce to those obtained in GR
when $f(G,T)=0$.

\section*{Appendix A}
\renewcommand{\theequation}{A\arabic{equation}}
\setcounter{equation}{0}
The modified terms in $f(G,T)$ gravity
appearing in the field equations are
\begin{eqnarray}\nonumber
T^{(GT)}_{00}&=&\frac{1}{8\pi}\left[\left(\mu+P\right)A^2f_{T}-\frac{A^2}{2}f+\left(\frac{8\dot{B}\dot{C}\ddot{C}}{A^2B
C^2}-\frac{16A'\dot C\dot
C'}{AB^2C^2}+\frac{8AC'A'C''}{B^4C^2}\right.\right.\\\nonumber
&+&\left.\left. \frac{4\ddot B}{BC^2}+\frac{8A'\dot B\dot
CC'}{AB^3C^2}+\frac{8\dot A\dot CC''}{AB^2C^2}-\frac{8\dot AB'C'\dot
C}{AB^3C^2}-\frac{12\dot A\dot B\dot C^2}{A^3BC^2}+\frac{4\dot
C^2A'}{AC^2}\right.\right.\\\nonumber
&\times&\left.\left.\frac{B'}{B^3}-\frac{12AA'B'C'^2}
{B^5C^2}+\frac{4C'^2\dot A\dot
B}{AB^3C^2}+\frac{4AA''C'^2}{B^4C^2}+\frac{4\dot C^2\ddot B
}{A^2BC^2}+\frac{8\dot C'^2}{B^2C^2}\right.\right.\\\nonumber
&+&\left.\left.\frac{8B'C'\ddot C}{B^3C^2}-\frac{16\dot B\dot
C'C'}{C^2B^3}-\frac{4\dot A\dot B}{AC^2B}-\frac{4\dot
C^2A''}{B^2AC^2}+\frac{8A'^2\dot C^2}{B^2A^2C^2}+\frac{8C'^2\dot
B^2}{C^2B^4}\right.\right.\\\nonumber &-&\left.\left.\frac{8\ddot
CC''}{C^2B^2}-\frac{4AA''}{C^2B^2}+\frac{B'AA'}{B^3C^2}
-\frac{4C'^2\ddot B}{B^3C^2}\right)f_{G}+\left(\frac{8\dot
CC''}{B^2C^2}-\frac{8B'C'\dot C}{B^3C^2}\right.\right.\\\nonumber
&-&\left.\left.\frac{4\dot B}{BC^2}-\frac{12\dot B\dot
C^2}{A^2BC^2}+\frac{4\dot BC'^2}{B^3C^2}\right)\dot
f_{G}+\left(\frac{8\dot B \dot CC'}{B^3C^2}-\frac{4B'\dot
C^2}{B^3C^2}-\frac{4A^2B'}{B^3C^2}\right.\right.
\\\nonumber&+&\left.\left.\frac{12A^2B'C'^2}{B^5C^2}
-\frac{8A^2C'C''}{B^4C^2}\right)f'_{G} +\left(\frac{4\dot
C^2}{B^2C^2}+\frac{4A^2}{B^2C^2}-\frac{4A^2C'^2}{B^4C^2}\right)\right.\\\label{100}
&\times&\left.f''_{G}\right],
\\\nonumber
T^{(GT)}_{01}&=&\frac{1}{8\pi}\left[-qABf_{T}+\left(\frac{10A'B'C'\dot
B}{AB^4C}-\frac{10\dot A\dot B\dot CA'}{A^4BC}-\frac{8\dot A\dot
B\dot C'}{A^3BC^2}-\frac{8A' }{AB^3}\right.\right.\\\nonumber
&\times&\left.\left.\frac{\dot BC'^2}{C^2}-\frac{10\dot A\dot
B^2C'}{A^3B^2C}-\frac{8A'^2C'\dot C}{A^2B^2C^2}+\frac{10A'^2B'\dot
C}{A^2B^3C}+\frac{\dot B\ddot CC'}{A^2BC^2}+\frac{10\dot B \ddot
B}{A^2B^2}\right.\right.\\\nonumber
&\times&\left.\left.\frac{C'}{C}+\frac{10A'\ddot B\dot
C}{A^3BC}-\frac{10A''A'\dot C}{A^2B^2C}-\frac{10A''C'\dot
B}{AB^3C}+\frac{\dot A\dot B\dot
C'}{A^3BC}+\frac{8A'C'}{AB^2}-10\right.\right.\\\nonumber
&-&\left.\left.\frac{A'B'\dot C'\dot C'}{AB^3CC^2}-\frac{8\dot
AA'\dot C^2}{A^4C^2}+\frac{8A'\dot C\ddot C}{A^3C^2}+\frac{8\dot
A\dot C\dot C'}{A^3C^2}-\frac{10\ddot B\dot
C'}{A^2BC}+\frac{10A''\dot
C'}{AB^2C}\right.\right.\\\nonumber&-&\left.\left.\frac{8\ddot C\dot
C'}{A^2C^2}\right)f_{G}+\left(-\frac{8\dot A\dot
BA'}{A^4B}-\frac{4A'}{AC^2}+\frac{8\dot C\dot
C'}{A^2C^2}+\frac{4A'C'^2}{AB^2C^2}+\frac{8A'^2B'}{A^2B^3}\right.\right.\\\nonumber&-&\left.\left.\frac{12A'\dot
C^2}{A^3C^2}-\frac{8A'A''}{A^2B^2}+\frac{8A'\ddot
B}{A^3B}\right)\dot f_{G}\left(\frac{8A'C'\dot
C}{AB^2C^2}+\frac{8A'B'\dot B}{AB^4}-\frac{4\dot B\dot
C^2}{A^2BC^2}\right.\right.\\\nonumber&-&\left.\left.\frac{8\dot
A\dot B^2}{A^3B^2}+\frac{12\dot BC'^2}{B^3C^2}+\frac{8\ddot B\dot
B}{A^2B^2}-\frac{8A''\dot B}{AB^3}-\frac{8C'\dot
C'}{B^2C^2}-\frac{4\dot B}{BC^2}\right)f'_{G}+\left(4\right.\right.
\\\label{100a}&\times&\left.\left.\frac{\dot
C^2}{A^2C^2}+\frac{4}{C^2}-\frac{4C'^2}{B^2C^2}+\frac{8A''}{AB^2}-\frac{8\ddot
B}{A^2B}-\frac{8A'B'}{AB^3}+\frac{8\dot A\dot B}{A^3B}\right)\dot
f'_{G}\right],\\\nonumber
T^{(GT)}_{11}&=&\frac{1}{8\pi}\left[\frac{2}{3}\Pi B^2
f_{T}+\frac{B^2}{2}f+\left(\frac{32\dot BC' \dot
C'^2}{A^2BC^2}-\frac{8A'C'C''}{AB^2C^2}-\frac{8B\dot B\dot C\ddot
C}{A^2C^4}-8\right.\right.\\\nonumber
&\times&\left.\left.\frac{B'\ddot CC'}{BC^2A^2}+\frac{\ddot
BC'^2}{A^2BC^2}-\frac{4A''C'^2}{AB^2C^2}-\frac{16\dot
C'^2}{A^2C^2}-\frac{8\dot A\dot CC''}{A^3C^2}+\frac{32A'\dot C\dot
C'}{A^3C^2}\right.\right.\\\nonumber &-&\left.\left.\frac{4B\ddot
B\dot C^2}{A^4C^2}-\frac{\dot B^2C'^2}{A^2B^2C^2}+\frac{8\ddot
CC''}{A^2C^2}+\frac{4B\dot B\dot
A}{A^3C^2}-\frac{4A'B'}{ABC^2}+\frac{4A''\dot
C^2}{A^3C^2}-24\right.\right.\\\nonumber
&\times&\left.\left.\frac{A'\dot B\dot CC'}{A^3BC^2}+\frac{8\dot
A\dot CB'C'}{A^3BC^2}-\frac{4B\ddot
B}{A^2C^2}+\frac{4A''}{AC^2}-\frac{16A'^2\dot
C^2}{A^4C^2}+\frac{12A'B'C'^2}{AB^3C^2}\right.\right.\\\nonumber
&-&\left.\left.\frac{4A'B'\dot C^2}{A^3BC^2}-\frac{4\dot A\dot
BC'^2}{A^3BC^2}\right)f_{G}+\left(\frac{8B^2\dot C\ddot
C}{A^4C^2}-\frac{12B^2\dot A\dot C^2}{A^5C^2}+\frac{4\dot
AC'^2}{A^3C^2}\right.\right.\\\nonumber &-&\left.\left.\frac{8A'\dot
CC'}{A^3C^2}-\frac{4B^2\dot A}{A^3C^2}\right)\dot
f_{G}+\left(\frac{4B^2\dot
C^2}{A^4C^2}\frac{4B^2}{A^2C^2}-\frac{4C'^2}{A^2C^2}\right)\ddot
f_{G}+\left(\frac{8\dot C}{A^3}\right.\right.\\\label{100b}
&\times&\left.\left.\frac{\dot AC'}{C^2}-\frac{4\dot
A'C^2}{A^3C^2}-\frac{4A'}{AC^2}+\frac{12C'^2A'}{AB^2C^2}-\frac{\ddot
CC'}{A^2C^2}\right)f'_{G}\right]\\\nonumber
T^{(GT)}_{22}&=&\frac{1}{8\pi}\left[\frac{ -C^2 }{3}\Pi
f_{T}+\frac{C^2}{2}f+\left(\frac{-4\dot
C^2}{A^2B^2}+\frac{4A''}{AB^2}-\frac{4\ddot B}{A^2B}+\frac{8\dot
A\dot
CB'C'}{A^3B^3}-4\right.\right.\\\nonumber&\times&\left.\left.\frac{\dot
AC'^2\dot B}{B^3A^3}-\frac{4A'\dot C^2B'}{B^3A^3}+\frac{12\dot B\dot
A\dot
C^2}{BA^5}+\frac{12A'B'C'^2}{AB^5}-\frac{8A'C''C'}{B^4A}-\frac{8\dot
A\dot
C}{A^3}\right.\right.\\\nonumber&\times&\left.\left.\frac{C''}{B^2}-\frac{4A'B'}{AB^3}--\frac{8C'B'\ddot
C}{B^3A^2}\frac{8\dot B\dot C \ddot C}{A^4B}+\frac{8A'\dot C\dot
C'}{A^3B^2}+\frac{4\dot A\dot
B}{A^3B}-\frac{4A''C'^2}{AB^4}\right.\right.\\\nonumber&-&\left.\left.\frac{4A'^2
\dot C^2}{A^4B^2}-\frac{4\ddot B \dot C^2}{A^4 B}+\frac{4\ddot
BC'^2}{A^2B^3}+\frac{4A''\dot C^2}{A^3B^2}+\frac{8\ddot
CC''}{A^2B^2}-\frac{4\dot
B^2C'^2}{A^2B^4}\right)f_{G}+\right.\\\nonumber&+&\left.\left(\frac{4\ddot
BC\dot C}{A^4B}-\frac{12\dot A\dot B \dot CC}{A^5B}-\frac{4\dot AB'C
C'}{A^3B^3}+\frac{8A'C'\dot B}{A^3B^3}+\frac{4A'B'\dot
CC}{A^3B^3}-\frac{4}{A^3}\right.\right.\\\nonumber&\times&\left.\left.\frac{A''\dot
CC}{B^2}+\frac{4\dot B\ddot CC }{A^4B}-\frac{12A'C\dot
C'}{A^3B^2}+\frac{4\dot ACC''}{A^3B^2}+\frac{12C\dot
CA'^2}{A^4B^2}\right)\dot
f_{G}+\left(8\right.\right.\\\nonumber&\times&\left.\left.\frac{A'\dot
B\dot CC}{A^3B^3}+\frac{4B'C\ddot C}{B^3A^2}+\frac{4C\dot A\dot
BC'}{A^3B^3}-\frac{12A'B'C'C}{AB^5}-\frac{4B'\dot A\dot
CC}{B^3A^3}+\frac{4}{B^4}\right.\right.\\\nonumber&\times&\left.\left.\frac{A''CC'}{A}+\frac{8\dot
BC'\dot C'}{A^2B^3}-\frac{4\ddot BC'C}{B^3A^2}-\frac{12\dot BC\dot
C'}{B^3A^2}+\frac{4A'CC''}{B^4A}+\frac{12A'^2}{A^2}\right.\right.\\\nonumber&\times&\left.\left.\frac{C\dot
C}{B^4}\right)f'_{G}+\left(\frac{4\dot A\dot
CC}{A^3B^2}+\frac{4A'C'C}{AB^4}-\frac{4\ddot
CC}{A^2B^2}\right)f''_{G}+\left(\frac{12C\dot
C'}{A^2B^2}-\frac{12C}{A^2}\right.\right.\\\label{100c}&\times&\left.\left.\frac{\dot
BC'}{B^3}-\frac{12A'C\dot C}{A^3B^2}\right)\dot
f'_{G}+\left(\frac{4\dot B\dot
CC}{A^4B}+\frac{4B'C'C}{A^2B^3}-\frac{4CC''}{A^2B^2}\right)\ddot
f_{G}\right].
\end{eqnarray}
The expressions $Z_1$ and $Z_2$ come out to be
\begin{eqnarray}\nonumber
Z_{1}&=&\frac{f_{T}}{k^{2}-f_{T}}\left[\left(\frac{\mu}{A}+
\frac{T^{00(GT)}}{A^3}+\frac{P}{A}\right)\left(\ln f_T
\right)^{.}+2\left(\frac{q}{B}-\frac{T^{01(GT)}}{AB^2}\right)+\frac{1}{2A}\right.\\\nonumber
&\times&\left.\left(\mu+3P\right)^.+\left(\frac{q}{B}-\frac{T^{01(GT)}}{AB^2}\right)
(\ln f_T)'
+\left(\frac{2\mu}{A}+\frac{2T^{00(GT)}}{A^3}+\frac{P}{A}\right)^.\right],\\\label{100d}\\\nonumber
Z_{2}&=&\frac{f_{T}}{k^{2}-f_{T}}\left[\left(\frac{-P_r}{B}+\frac{T^{11(GT)}}{B^3}
+\frac{P}{B}\right)\left(\ln f_T
\right)'+2\left(\frac{-q}{A}+\frac{T^{01(GT)}}{A^2B}\right)-\frac{1}{2}\right.\\\nonumber
&\times&\left.\frac{\left(\mu+3P\right)'}{B}+\left(\frac{-2P_r}{B}-\frac{2T^{11(GT)}}{B^3}
+\frac{P}{B}\right)'+\left(\frac{-q}{A}+\frac{T^{01(GT)}}{A^2B}\right)\right.\\\label{100e}&\times&\left.(\ln
f_T)^.\right].
\end{eqnarray}

\section*{Appendix B}
\renewcommand{\theequation}{B\arabic{equation}}
\setcounter{equation}{0}
The correction terms in the structure
scalars are evaluated as
\begin{eqnarray}\nonumber
M^{(GT)}&=& 2\left[R_{\mu\beta}R^{\mu}_{\alpha}f_{G}+
R^{\mu\nu}R_{\mu\beta\nu\alpha}f_{G}-\frac{1}{2}RR_{\alpha\beta}f_{G}-\frac{1}{2}R_{\beta\mu\nu
n}R^{\mu\nu n}_{\alpha}f_{G}\right.\\\nonumber&+&\left.\frac{1}{2}R
\nabla_{\alpha}\nabla_{\beta}f_{G}+R_{\alpha\beta}\Box
f_{G}-R^{\mu}_{\alpha}\nabla_{\beta}\nabla_{\mu}f_{G}-R^{\mu}_{\beta}\nabla_{\alpha}\nabla_{\mu}f_{G}\right.\\\nonumber
&-&\left.R_{\mu\beta
\nu\alpha}\nabla^{\mu}\nabla^{\nu}f_{G}\right]g^{\alpha\beta}+2\left[-R_{\mu\delta}R^{\mu}_{\alpha}f_{G}-
R^{\mu\nu}R_{\mu\delta
\nu\alpha}f_{G}+\frac{1}{2}RR_{\alpha\delta}f_{G}\right.\\\nonumber&+&\left.\frac{1}{2}R_{\delta
\mu\nu n}R^{\mu\nu n}_{\alpha}f_{G}-R_{\delta\alpha}\Box
f_{G}+R_{\mu\delta\nu\alpha}\nabla^{\mu}\nabla^{\nu}f_{G}+R^{\mu}_{\alpha}\nabla_{\delta}\nabla_{\mu}f_{G}\right.\\\nonumber&-&\left.\frac{1}{2}R
\nabla_{\alpha}\nabla_{\delta}f_{G}+R^{\mu}_{\delta}\nabla_{\alpha}\nabla_{\mu}
f_{G}\right]V_{\beta}V^{\delta}g^{\alpha\beta}+2\left[-R_{\mu\beta}R^{\mu\gamma}f_{G}\right.\\\nonumber&+&\left.
\frac{1}{2}R_{\beta\mu\nu n}R^{\mu\nu
n\gamma}f_{G}-+\frac{1}{2}RR^{\gamma}_{\beta}f_{G}
R^{\mu\nu}R^{\gamma}_{\mu\beta
\nu}f_{G}+R^{\mu\gamma}\nabla_{\beta}\nabla_{\mu}f_{G}\right.\\\nonumber&-&\left.\frac{1}{2}R
\nabla^{\gamma}\nabla_{\beta}f_{G}+R^{\gamma}_{\mu\beta
\nu}\nabla^{\mu}\nabla^{\nu}f_{G}-R^{\gamma}_{\beta}\Box
f_{G}+R^{\mu}_{\beta}\nabla^{\gamma}\nabla_{\mu}f_{G}\right]V_{\alpha}V_{\gamma}g^{\alpha\beta}\\\nonumber
&+&2\left[R_{\mu\delta}R^{\mu\gamma}f_{G}+
R^{\mu\nu}R^{\gamma}_{\mu\delta \nu}f_{G}-\frac{1}{2}R_{\delta
\mu\nu n}R^{\mu\nu n \gamma}f_{G}+\frac{1}{2}R
\nabla^{\gamma}\nabla_{\delta}f_{G}\right.\\\nonumber&+&\left.R^{\gamma}_{\delta}\Box
f_{G}-R^{\mu\gamma}\nabla_{\delta}\nabla_{\mu}f_{G}
-R^{\mu}_{\delta}\nabla^{\gamma}\nabla_{\mu}f_{G}-R^{\gamma}_{\mu\delta
\nu}\nabla^{\mu}\nabla^{\nu}f_{G}-\frac{1}{2}RR^{\gamma}_{\delta}f_{G}\right]
\\\nonumber &\times&g_{\alpha\beta}V_{\gamma}V^{\delta}g^{\alpha\beta}-\left[4R^{\mu m}R_{\mu m}f_{G}+4R^{\mu\nu}R^{m}_{\mu m \nu}f_{G}
-2R^{2}f_{G}\right.\\\nonumber&-&\left.2R^{l}_{\mu \nu n}R^{\mu \nu
n}_{l}f_{G}-4R\Box f_{G}+16R^{\mu\nu}\nabla_{\mu}\nabla_{\nu}f_{G}
-4R^{\mu
m}\nabla_{m}\nabla_{\mu}f_{G}\right.\\\nonumber&-&\left.4R^{\nu
l}\nabla_{l}\nabla_{\nu}f_{G}-4R^{m}_{\mu m
\nu}\nabla^{\mu}\nabla^{\nu}f_{G}\right]-
\frac{1}{2}f+12R^{\mu\nu}\nabla_{\mu}\nabla_{\nu}f_{G}-6R\Box f_{G},
\end{eqnarray}
\begin{eqnarray}\nonumber
J^{(GT)}_{(\alpha\beta)} &=& \left[2R_{\mu
d}R^{\mu}_{c}f_{G}+2R^{\mu\nu}R_{\mu d \nu
c}f_{G}-RR_{cd}f_{G}-R_{d\mu\nu n}R^{\mu\nu n}_{c}f_{G}+2R_{cd}\Box
f_{G}\right.\\\nonumber&+&\left.R\nabla_{c}\nabla_{d}f_{G}-2R^{\mu}_{c}\nabla_{d}\nabla_{\mu}f_{G}-2R^{\mu}_{d}\nabla_{c}\nabla_{\mu}f_{G}
-2R_{\mu d\nu c}\nabla^{\mu}\nabla^{\nu}f_{G}\right]
h^{c}_{\alpha}h^{d}_{\beta}\\\nonumber&+&2\left[R_{\mu\delta}R^{\mu\gamma}f_{G}+
R^{\mu\nu}R^{\gamma}_{\mu\delta
\nu}f_{G}-\frac{1}{2}RR^{\gamma}_{\delta}f_{G}-\frac{1}{2}R_{\delta
\mu\nu n}R^{\mu\nu n
\gamma}f_{G}\right.\\\nonumber&+&\left.R^{\gamma}_{\delta}\Box
f_{G}+\frac{1}{2}R
\nabla^{\gamma}\nabla_{\delta}f_{G}-R^{\mu\gamma}\nabla_{\delta}\nabla_{\mu}f_{G}
-R^{\mu}_{\delta}\nabla^{\gamma}\nabla_{\mu}f_{G}\right.\\\nonumber&-&\left.R^{\gamma}_{\mu\delta
\nu}\nabla^{\mu}\nabla^{\nu}f_{G}\right]h_{\alpha\beta}V_{\gamma}V^{\delta}-2\left[R_{\mu\beta}R^{\mu}_{\alpha}f_{G}+
R^{\mu\nu}R_{\mu\beta
\nu\alpha}f_{G}-\frac{1}{2}RR_{\alpha\beta}f_{G}\right.\\\nonumber&-&\left.\frac{1}{2}R_{\beta
\mu\nu n}R^{\mu\nu n}_{\alpha}f_{G}+R_{\alpha\beta}\Box
f_{G}+\frac{1}{2}R
\nabla_{\alpha}\nabla_{\beta}f_{G}-R^{\mu}_{\alpha}\nabla_{\beta}\nabla_{\mu}f_{G}\right.\\\nonumber&-&\left.
R^{\mu}_{\beta}\nabla_{\alpha}\nabla_{\mu}f_{G}-R_{\mu\beta
\nu\alpha}\nabla^{\mu}\nabla^{\nu}f_{G}\right]-2\left[-R_{\mu\delta}R^{\mu}_{\alpha}f_{G}-
R^{\mu\nu}R_{\mu\delta
\nu\alpha}f_{G}\right.\\\nonumber&+&\left.\frac{1}{2}RR_{\alpha\delta}f_{G}+\frac{1}{2}R_{\delta
\mu\nu n}R^{\mu\nu n}_{\alpha}f_{G}-R_{\delta\alpha}\Box
f_{G}+R_{\mu\delta
\nu\alpha}\nabla^{\mu}\nabla^{\nu}f_{G}\right.\\\nonumber&+&\left.R^{\mu}_{\alpha}\nabla_{\delta}\nabla_{\mu}f_{G}
+R^{\mu}_{\delta}\nabla_{\alpha}\nabla_{\mu}f_{G}-\frac{1}{2}R
\nabla_{\alpha}\nabla_{\delta}f_{G}\right]V_{\beta}V^{\delta}-2\left[-R_{\mu\beta}R^{\mu\gamma}f_{G}\right.
\\\nonumber&-&\left.R^{\mu\nu}R^{\gamma}_{\mu\beta
\nu}f_{G}+\frac{1}{2}RR^{\gamma}_{\beta}f_{G}+\frac{1}{2}R_{\beta
\mu\nu n}R^{\mu\nu n\gamma}f_{G}-R^{\gamma}_{\beta}\Box
f_{G}\right.\\\nonumber&+&\left.R^{\mu\gamma}\nabla_{\beta}\nabla_{\mu}f_{G}+R^{\mu}_{\beta}\nabla^{\gamma}\nabla_{\mu}f_{G}+R^{\gamma}_{\mu\beta
\nu}\nabla^{\mu}\nabla^{\nu}f_{G}-\frac{1}{2}R
\nabla^{\gamma}\nabla_{\beta}f_{G}\right]V_{\alpha}V_{\gamma}\\\nonumber&-&2\left[R_{\mu\delta}R^{\mu\gamma}f_{G}+
R^{\mu\nu}R^{\gamma}_{\mu\delta
\nu}f_{G}-\frac{1}{2}RR^{\gamma}_{\delta}f_{G}-\frac{1}{2}R_{\delta
\mu\nu n}R^{\mu\nu n
\gamma}f_{G}\right.\\\nonumber&+&\left.R^{\gamma}_{\delta}\Box
f_{G}+\frac{1}{2}R
\nabla^{\gamma}\nabla_{\delta}f_{G}-R^{\mu\gamma}\nabla_{\delta}\nabla_{\mu}f_{G}
-R^{\mu}_{\delta}\nabla^{\gamma}\nabla_{\mu}f_{G}\right.\\\nonumber&-&\left.R^{\gamma}_{\mu\delta
\nu}\nabla^{\mu}\nabla^{\nu}f_{G}\right]V_{\gamma}V^{\delta}g_{\alpha\beta},
\end{eqnarray}
\begin{eqnarray}\nonumber
Q^{(GT)} &=&\left[\frac{1}{2}R_{\mu\epsilon}R^{\mu
p}f_{G}+\frac{1}{2}R^{\mu\nu}R^{p}_{\mu\epsilon
\nu}f_{G}-\frac{1}{4}RR^{p}_{\epsilon}f_{G}-\frac{1}{4}R_{\epsilon
\mu\nu n}R^{\mu\nu n p}f_{G}\right.\\\nonumber&+&\left.\frac{1}{2}
R^{p}_{\epsilon}\Box
f_{G}+\frac{1}{4}R\nabla^{p}\nabla_{\epsilon}f_{G}-\frac{1}{4}R^{\mu
p}\nabla_{\epsilon}\nabla_{\mu}f_{G}-\frac{1}{2}R^{\mu}_{\epsilon}\nabla^{p}\nabla_{\mu}
f_{G}\right.\\\nonumber&-&\left.\frac{1}{2}R^{p}_{\mu\epsilon
\nu}\nabla^{\mu}\nabla^{\nu}f_{G}\right]g^{\alpha\beta}\epsilon_{p
\delta\beta}\epsilon^{\epsilon\delta}_{\alpha}+\left[-\frac{1}{2}R_{\mu\delta}R^{\mu
p}f_{G}-\frac{1}{2}R^{\mu\nu}R^{p}_{\mu\delta \nu}
f_{G}\right.\\\nonumber&+&\left.\frac{1}{4}RR^{p}_{\delta}f_{G}+\frac{1}{4}R_{\delta
\mu\nu n}R^{\mu\nu np}f_{G}-\frac{1}{2}R^{p}_{\delta}\Box
f_{G}-\frac{1}{4}R\nabla^{p}\nabla_{\delta}f_{G}\right.\\\nonumber&+&\left.\frac{1}{4}R^{\mu
p}\nabla_{\delta}\nabla_{\mu}f_{G}+\frac{1}{2}R^{\mu}_{\delta}\nabla^{p}\nabla_{\mu}f_{G}
+\frac{1}{2}R^{p}_{\mu\delta
\nu}\nabla^{\mu}\nabla^{\nu}f_{G}\right]g^{\alpha\beta}\epsilon_{p\epsilon
\beta}\epsilon^{\epsilon\delta}_{\alpha}\\\nonumber&+&\left[-\frac{1}{2}R_{\mu\epsilon}R^{\mu\gamma}f_{G}-\frac{1}{2}R^{\mu\nu}R^{\gamma}_{\mu\epsilon
\nu}f_{G}
+\frac{1}{4}RR^{\gamma}_{\epsilon}f_{G}+\frac{1}{4}R_{\epsilon
\mu\nu n}R^{\mu\nu
n\gamma}f_{G}\right.\\\nonumber&-&\left.\frac{1}{2}R^{\gamma}_{\epsilon}\Box
f_{G}-\frac{1}{4}R\nabla^{\gamma}\nabla_{\epsilon}f_{G}+\frac{1}{4}R^{\mu\gamma}\nabla_{\epsilon}\nabla_{\mu}f_{G}
+\frac{1}{2}R^{\mu}_{\epsilon}\nabla^{\gamma}\nabla_{\mu}
f_{G}\right.\\\nonumber&+&\left.\frac{1}{2}R^{\gamma}_{\mu\epsilon
\nu}\nabla^{\mu}
\nabla^{\nu}f_{G}\right]g^{\alpha\beta}\epsilon_{\delta\gamma\beta}\epsilon^{\epsilon\delta}_{\alpha}+\left[\frac{1}{2}R_{\mu\delta}R^{\mu\gamma}f_{G}
+\frac{1}{2}R^{\mu\nu}R^{\gamma}_{\mu\delta
\nu}f_{G}\right.\\\nonumber&-&\left.\frac{1}{4}RR^{\gamma}_{\delta}f_{G}-\frac{1}{4}R_{\delta
\mu\nu n}R^{\mu\nu n\gamma}f_{G}+\frac{1}{2}R^{\gamma}_{\delta}\Box
f_{G}+\frac{1}{4}R\nabla^{\gamma}\nabla_{\delta}f_{G}\right.\\\nonumber&-&\left.\frac{1}{4}R^{\mu\gamma}\nabla_{\delta}\nabla_{\mu}f_{G}
-\frac{1}{2}R^{\mu}_{\delta}\nabla^{\gamma}\nabla_{\mu}f_{G}-\frac{1}{2}R^{\gamma}_{\mu\delta
\nu}\nabla^{\mu}\nabla^{\nu}f_{G}\right]g^{\alpha\beta}\epsilon_{\epsilon\gamma\beta}\epsilon^{\epsilon\delta}_{\alpha}\\\nonumber&-&
12R^{\mu\nu}\nabla_{\mu}\nabla_{\nu}f_{G} +6R\Box
f_{G}+\left[\left(\mu+P\right)f_{T}+4R^{\mu\nu}R_{\mu\nu}f_{G}\right.\\\nonumber&+&\left.4R^{\mu\nu}R^{m}_{\mu\nu
m}f_{G}-2R^{2}f_{G}-2R^{l}_{\mu\nu n}R^{\mu\nu n}_{l}f_{G}-4R\Box
f_{G}\right.\\\nonumber&+&\left.16R^{\mu\nu}\nabla_{\mu}\nabla_{\nu}f_{G}-4R^{\mu
m}\nabla_{m}\nabla_{\mu}f_{G} -4R^{\mu
l}\nabla_{l}\nabla_{\mu}f_{G}\right.\\\nonumber&-&\left.4R^{m}_{\mu
m \nu}\nabla^{\mu}\nabla^{\nu}f_{G}\right]+\frac{1}{2}f.
 \end{eqnarray}

\end{document}